\documentclass[%
 reprint,
 amsmath,amssymb,
 aip,
]{revtex4-2}
\usepackage[most]{tcolorbox}
\usepackage{tikz}
\usepackage{graphicx}
\usepackage{dcolumn}
\usepackage{bm}
\usepackage{relsize}
\usepackage[normalem]{ulem}
\usepackage{hyperref}
\usepackage{xcolor}

\begin{document}


\title{Interfaces as transport barriers in two-dimensional Cahn-Hilliard-Navier-Stokes turbulence}

\author{Nadia Bihari Padhan}
 \email{nadia@iisc.ac.in}
\author{Rahul Pandit}%
 \email{rahul@iisc.ac.in}
\affiliation{%
 Department of Physics, Indian Institute of Science, Bangalore 560012, India. 
}%

\date{\today}

\begin{abstract}
 We investigate the role of interfaces as transport barriers in binary-fluid turbulence by employing Lagrangian tracer particles. The Cahn-Hilliard-Navier-Stokes (CHNS) system of partial differential equations provides a natural theoretical framework for our investigations. For specificity, we utilize the two-dimensional (2D) CHNS system. We capture efficiently interfaces and their fluctuations in 2D binary-fluid turbulence by using extensive pseudospectral direct numerical simulations (DNSs) of the 2D CHNS equations. We begin with $n$ tracers within a droplet of one phase and examine their dispersal into the second phase. The tracers remain within the droplet for a long time before emerging from it, so interfaces act as transport barriers in binary-fluid turbulence. We show that the fraction of the number of particles inside the droplet decays exponentially and is characterized by a decay time $\tau_{\xi}\sim R_0^{3/2}$ that increases with $R_0$, the radius of the initially circular droplet. Furthermore, we demonstrate that the average first-passage time $\langle \tau \rangle$ for tracers inside a droplet is orders of magnitude larger than it is for transport out of a hypothetical circle with the same radius as the initially circular droplet. We examine the roles of the Okubo-Weiss parameter $\Lambda$, the fluctuations of the droplet perimeter, and the probability distribution function of $\cos(\theta)$, with $\theta$ the angle between the fluid velocity and the normal to a droplet interface, in trapping tracers inside droplets. We mention possible generalisations of our study. 
\end{abstract}

\maketitle
\newpage

\section{Introduction}

Droplet dynamics in multi-phase flows, both laminar and turbulent, poses considerable challenges for experimental, numerical, and theoretical studies: Not only must we resolve the dynamics of the immiscible fluids in a multi-phase flow, but we must also track the spatiotemporal evolution of the fluctuating interfaces that separate the coexisting fluids [see, e.g., Refs.~\cite{scardovelli1999direct,rider1998reconstructing,roccon2023phase}].  In particular, we must monitor mass- and momentum-transfer across these interfaces whose deformations can affect the flow topology, which has been studied for droplet-laden turbulent flows in Refs.~\cite{dodd2019small, soligo2020effect} and for droplets in turbulent shear flows in Ref.~\cite{rosti2019droplets}. The interactions between particles and droplet interfaces in channel-flow turbulence has been reported in Ref.~\cite{hajisharifi2021particle}.\\

The Cahn-Hilliard-Navier-Stokes (CHNS) system of partial differential equations (PDEs) combine the Navier-Stokes equations for a Newtonian fluid~\cite{navier1823memoire,doering1995applied} and the Cahn-Hilliard equations for binary mixtures~\cite{cahn1958free,cahn1959free,cahn1961spinodal,hohenberg1977theory,onuki2002phase,puri2009kinetics,lee2014physical}. The CHNS system provides a natural framework for studying multi-phase flows~\cite{Jacqmin_1999,Magaletti_2013,Boyer_2006,mirjalilimodeling,roccon2023phase,celani2009phase,lee2013numerical,gibbon2016regularity,lee2015two}, in general, and droplet dynamics, in particular. For example, the CHNS system has been used to uncover the multifractality of droplet-perimeter fluctuations in both conventional~\cite{Pal_2016,palphdthesis} and active-fluid flows~\cite{padhan2023activity}; the spatiotemporal evolution of liquid-lens mergers has been studied by using a three-phase version of the CHNS PDEs in Ref.~\cite{padhan2023unveiling}. We show that the role of interfaces as transport barriers in binary-fluid turbulence can be addressed in detail by monitoring the advection of Lagrangian tracer particles by CHNS flows. Such tracers have been employed widely in various forms of turbulence, including fluid and MHD turbulence, bacterial turbulence, and quantum turbulence [see, e.g., Refs.~\cite{yeung2002lagrangian,sreenivasan2010lagrangian,biferale2006lagrangian,homann2007lagrangian,toschi2009lagrangian,pandit2009statistical,la2013lagrangian,singh2022lagrangian,kiran2023irreversibility}].\\

To illustrate how interfaces act as transport barriers, we carry out extensive direct numerical simulations (DNSs) of the two-dimensional (2D) CHNS equations with such Lagrangian tracers. Our DNSs lead to important insights that we summarise qualitatively below before we present the details of our work.
If we initiate our DNS with tracers inside an intially circular droplet, which is advected by a turbulent flow, the tracers remain entrapped within the droplet for a long time before emerging from it. This demonstrates clearly that interfaces act as transport barriers for such tracers in binary-fluid turbulence. We show that the fraction of the number of particles inside the droplet decays exponentially and is characterized by a decay time $\tau_{\xi}\sim R_0^{3/2}$ that increases with $R_0$, the radius of the initially circular droplet. We calculate the average first-passage time $\langle \tau \rangle$ for tracers in a droplet and show that it is at least two orders of magnitude larger than it is for transport out of a hypothetical circle with the same radius as the initially circular droplet. We also elucidate the roles of the Okubo-Weiss parameter $\Lambda$, the fluctuations of the droplet perimeter, and the probability distribution function of $\cos(\theta)$, with $\theta$ the angle between the fluid velocity and the normal to a droplet interface, in trapping tracers inside droplets.

The remainder of this paper is organised as follows. We introduce the CHNS model and the numerical methods that we use in Section~\ref{sec:model}. We give a detailed presentation of our results in Section~\ref{sec:results}.
We end with a discussion of the significance of our results in Section~\ref{sec:conclusions}.

\section{Cahn-Hilliard-Navier-Stokes Equations}
\label{sec:model}

We consider the incompressible CHNS system of equations that couple the fluid velocity $\bm u$ with a scalar order parameter $c$ that changes continuously between the two fluid phases A and B. The fluid dynamics of $\bm u$ and $c$ is governed by the following CHNS PDEs~\cite{Jacqmin_1999, Boyer_2006, Magaletti_2013, fan2016cascades, fan2018chns, chaikin1995principles, Pal_2016} [we restrict our discussion to two spatial dimensions (2D)]:
\begin{eqnarray}
    \partial_t c + (\bm u \cdot \nabla) c &=& M \nabla^2 \mu\,; \label{eq:phi}\\
    \mu = \left( \frac{\delta \mathcal F}{\delta c}\right) &=& -\frac{3}{2}\sigma\epsilon \nabla^2c + 24 \frac{\sigma}{\epsilon} c(1 - c)(1 - 2c)\,;\\
    \partial_t \omega + (\bm u \cdot \nabla) \omega &=& \nu \nabla^2 \omega -\alpha \omega \\ \nonumber &+& [\nabla \times (\mu \nabla c)]\cdot \hat e_z + f^\omega;\label{eq:omega}\\
    \nabla \cdot \bm u &=& 0\,; \quad \omega = (\nabla \times \bm u)\cdot \hat e_z\,;\label{eq:incom}
\end{eqnarray}
the parameters $\nu$, $\alpha$, and $M$ are the kinematic viscosity, bottom friction, and mobility, respectively; the fluid density $\rho$ is a constant that we set to $1$. In Eq.~(\ref{eq:omega}) we employ the vorticity-streamfunction ($\omega-\psi$) formulation with
\begin{equation}
    \bm u = \nabla \times (\psi \hat e_z) \;\; {\rm{and}} \;\;
    \psi = - \nabla^{-2}\omega\,.
    \label{eq:ompsi}
    \end{equation}
The chemical potential $\mu$ is derived from the variational derivative of the Landau-Ginzburg free-energy functional
    \begin{eqnarray}
    \mathcal F[c, \nabla c] = \int_{\Omega} \left[12 \frac{\sigma}{\epsilon}c^2(1 - c)^2 + \frac{3}{4} \sigma \epsilon |\nabla c|^2\right]d\Omega\,;\label{eq:functional}
    \end{eqnarray}
    $\sigma$ and $\epsilon$, which can be varied independently, are the surface tension and interface width, respectively. The first term is a double-well potential with minima at $c = 0$ and $1$, which correspond
    to A-rich and B-rich bulk phases in equilibrium; and the second term gives the free-energy penalty for interfaces between these phases. The order parameter $c$ varies smoothly across an interface; $c\simeq 0$ in phase A and $c \simeq 1$ in phase B. Note that the scalar field $c$ is not a passive scalar [unlike the scalar in Ref.~\cite{jain2023computational}]; $c$ affects the flow velocity $\bm u$ even as it is advected by $\bm u$. If a strictly positive $c$ is required, then the term $c^2(1 - c)^2$ in Eq.~(\ref{eq:functional}) should be replaced by $[c\ln(c) + (1-c)\ln(1-c)]$.
    
    We generate turbulence by using a Kolmogorov-type forcing~\cite{pandit2017overview} of the form 
    \begin{equation}
        f^{\omega} = f_0 k_f \cos(k_f x)\,;\label{eq:force}
    \end{equation}
    $f_0$ and $k_f$ are the forcing amplitude and forcing wave-number, respectively.

\subsection{Numerical methods}
\label{subsec:dns}

We carry out extensive direct numerical simulations (DNSs) of the coupled equations~(\ref{eq:phi})-(\ref{eq:incom}) by using pseudospectral methods~\cite{canuto2012spectral}, with periodic boundary conditions in a square box of size $(2\pi \times 2\pi)$ with $1024 \times 1024$ collocation points. Our DNSs employ the semi-implicit exponential time difference Runge-Kutta-2 (ETDRK-2) method~\cite{cox2002exponential} for time integration and the $1/2$-dealiasing scheme to remove aliasing errors. To resolve the interface well, we take three computational grid points at interfaces.

The spatiotemporal evolution of the fields in Eqs.~(\ref{eq:phi})-(\ref{eq:incom}) depend on the initial conditions (see below) and the non-dimensional control parameters given hereunder: 
\begin{eqnarray}
    {\rm{Reynolds\; number}}\; Re &=& \frac{L_0 U_0}{\nu}\,;\\ {\rm{Cahn\; number}}\; Cn &=& \frac{\epsilon}{L_0}\,; \\ 
        {\rm{Weber\; number}}\; We &=& \frac{L_0 U_0^2}{\sigma}\,;\\ {\rm{Peclet\; number}}\; Pe &=& \frac{L_0 U_0 \epsilon}{M \sigma}\,; \\
        {\text{ non-dimensional\; friction}}\; \alpha^{\prime} &=& \frac{\alpha L_0}{U_0}\,.
\label{eq:parameters}
\end{eqnarray}
We use the following characteristic velocity, length, and time scales: 
\begin{eqnarray}
U = \frac{f_0}{\nu k_f^2}\,;\;\; L_0 = \frac{2\pi}{k_f}\,;\;\; T_0 = \frac{\nu k_f}{f_0}\,.
\label{eq:scales}
\end{eqnarray}
For all our DNSs, we keep the following parameters fixed: $Cn = 0.01$, $Re = 1178$, $We = 30$, $Pe = 1445$, and $\alpha^{\prime} = 0.004$.  
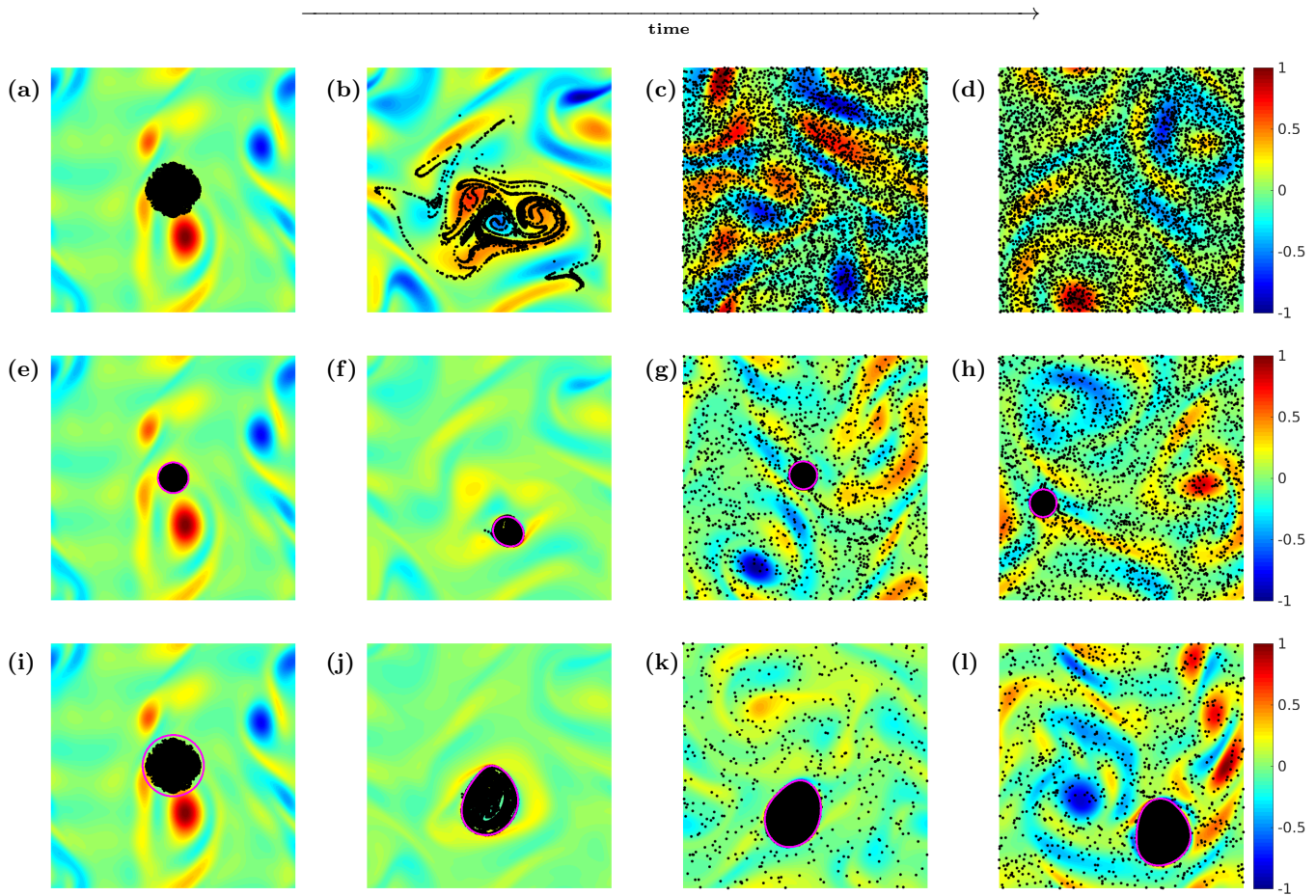
\begin{figure*}
{
\includegraphics[width=\textwidth]{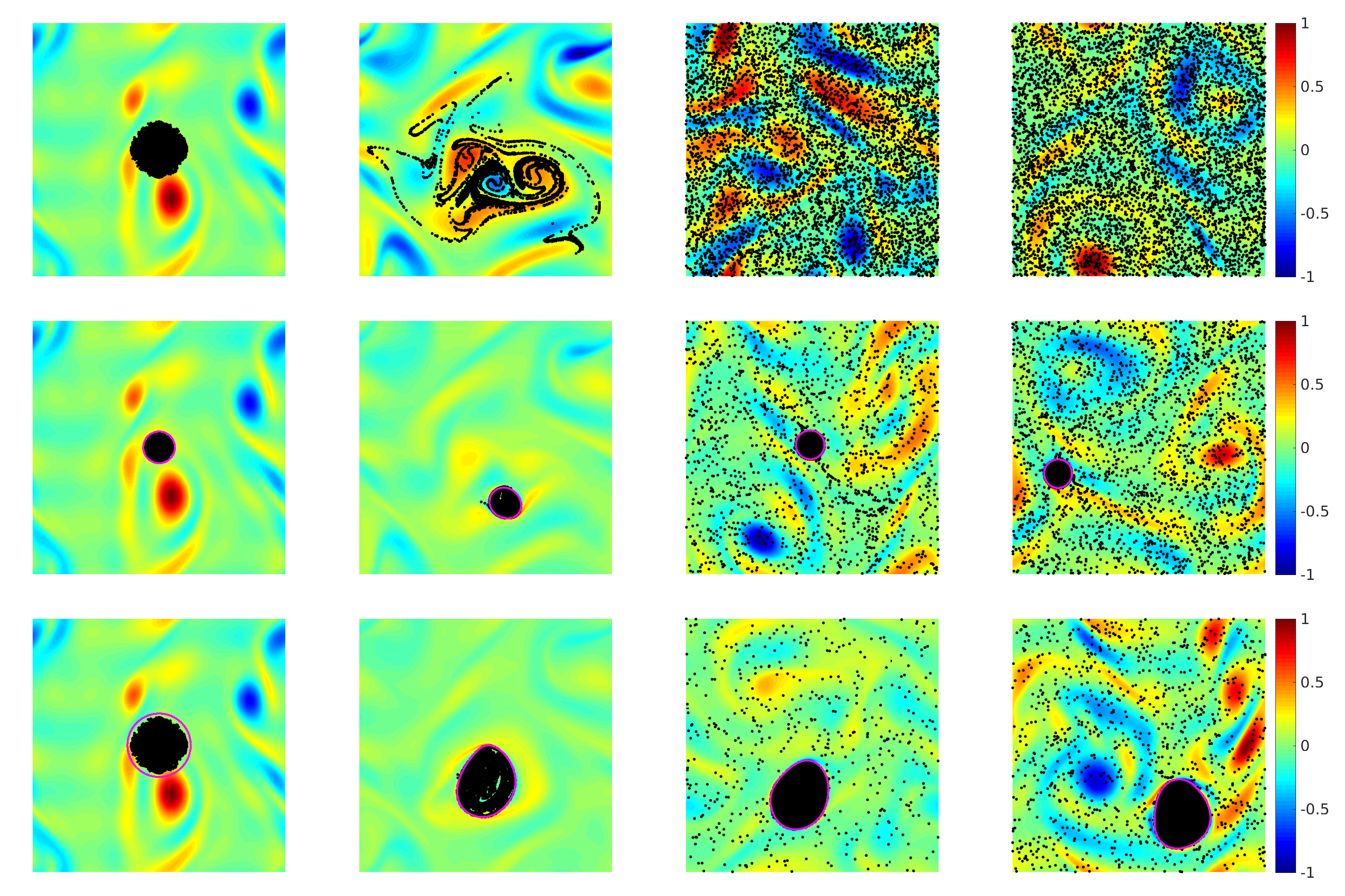}
\put(-400,350){\rm {$\xrightarrow[\textit{\bf{time}}]{\hspace*{10.0cm}}$}}
\put(-515,320){\rm {\bf(a)}}
\put(-390,320){\rm {\bf(b)}}
\put(-265,320){\rm {\bf(c)}}
\put(-145,320){\rm {\bf(d)}}
\put(-515,210){\rm {\bf(e)}}
\put(-390,210){\rm {\bf(f)}}
\put(-265,210){\rm {\bf(g)}}
\put(-145,210){\rm {\bf(h)}}
\put(-515,95){\rm {\bf(i)}}
\put(-390,95){\rm {\bf(j)}}
\put(-265,95){\rm {\bf(k)}}
\put(-145,95){\rm {\bf(l)}}
}

\caption{\label{fig:phi_tracers} Illustrative pseudocolor plots of $\omega$, with the $c = 0.5$ contour shown in magenta, at different representative simulation times (increasing from left to right): $t_1 = 0$, $t_2 = 15$, $t_3 = 150$, and $t_4 = 300$. The tracers are shown as black points for (a)-(d) droplet-free flow, (e)-(h) a droplet with initial radius $R_0/L_0 = 0.25$, and (i)-(l) a droplet with initial radius $R_0/L_0 = 0.5$; the
Weber number $\rm{We} = 30$. We use $n = 1024 \times 128 = 131072$ tracer particles in our DNSs but, for the figures here, we show the positions of only $6000$ particles. The vorticity $\omega$ is normalized by its maximum absolute value for ease of visualization. [For the full spatiotemporal evolution, see the videos V1-V3 in Appendix~\ref{sec:videos}.]}
\end{figure*}
\subsection{Lagrangian tracers}
\label{subsec:lagrangian}

To characterize the transport across a droplet's interface, we introduce $n$ Lagrangian tracer particles inside the droplet~\cite{lalescu2018tracer}. The temporal evolution of a tracer, with position $\bm r_0$ at time $t_0$, is given by 
\begin{eqnarray}
    \frac{d\bm r(t)}{dt} = \bm{v}(\bm r, t|\bm r_0, t_0) = \bm u(\bm r, t) \;, \label{eq:tracer}
\end{eqnarray}
where $\bm v$ is the Lagrangian velocity, which we obtain from the Eulerian velocity field $\bm u$~\cite{verma2020first,benzi2010inertial} [by using bilinear interpolation at off-grid points]. We use a first-order Euler scheme for the time integration of Eq.~(\ref{eq:tracer}).

\subsection{Initial conditions}
\label{subsec:init}
    We begin with an initially circular droplet of radius $R_0$ and with center at $(x_0, y_0) = (\pi, \pi)$:
    \begin{eqnarray}
    c(x, y, t=0) &=& 0.5\left[1 + \tanh{\left(\frac{R_0 - \sqrt{(x-x_0)^2 + (y-y_0)^2}}{\epsilon/2}\right)}\right] \,. \nonumber\\
     \label{eq:init}
    \end{eqnarray}
 This field $c(\bm x, 0)$ is used as an initial condition for Eqs.~(\ref{eq:phi})-(\ref{eq:incom}) in
 which the turbulent vorticity field $\omega$ is generated by the forcing term in Eq.~(\ref{eq:force}). The Reynolds number of the turbulent flow we use in our DNS is $Re = 1178$.

\section{Results}
\label{sec:results}
\begin{figure*}
{
\includegraphics[width=\textwidth]{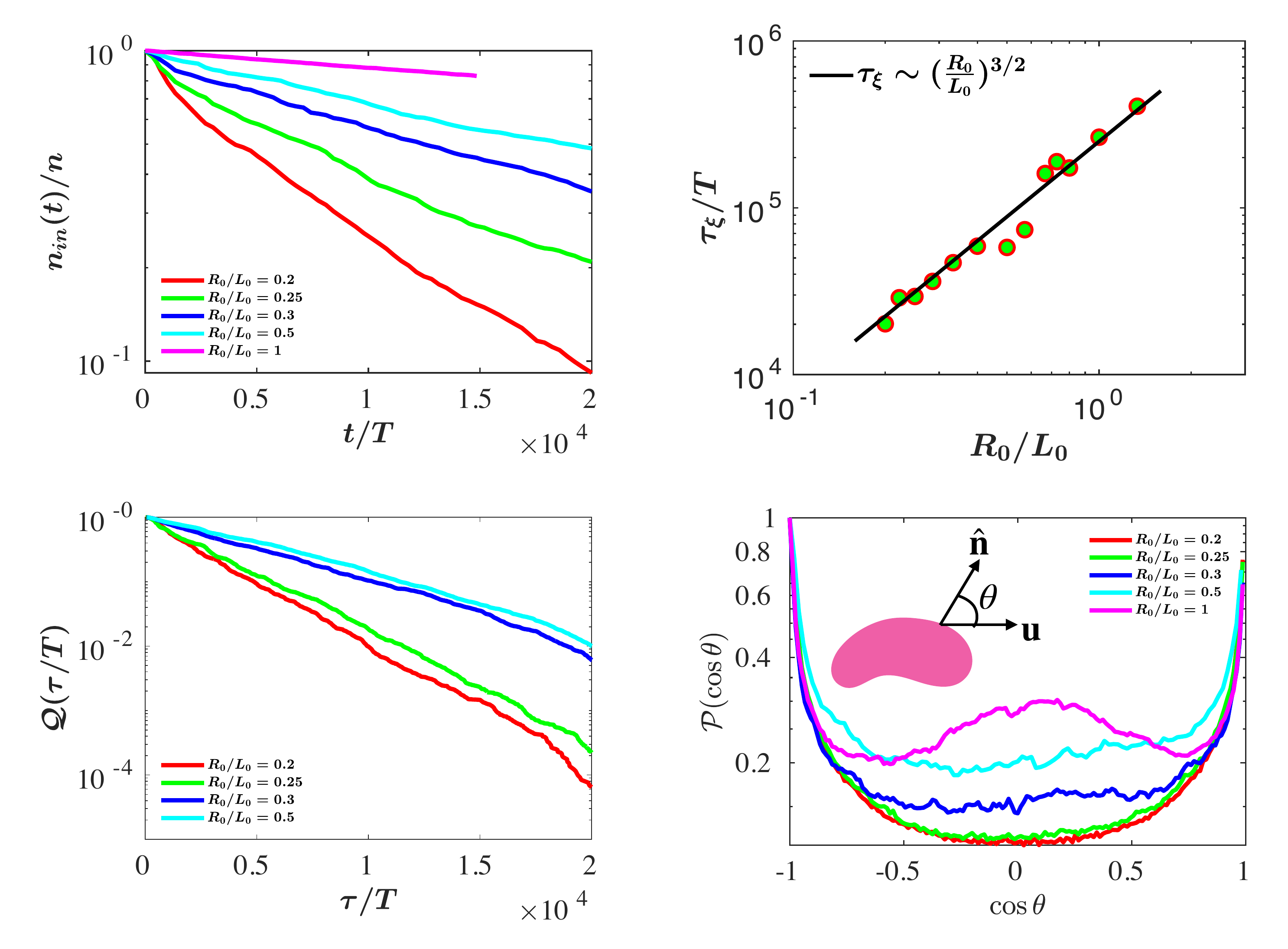}
\put(-500,350){\rm {\bf(a)}}
\put(-250,350){\rm {\bf(b)}}
\put(-500,160){\rm {\bf(c)}}
\put(-250,160){\rm {\bf(d)}}
}

\caption{\label{fig:n_tau} (a) Semi-log plots illustrating the exponential decay of the fraction of particles $n_{in}(t)/n$ inside the droplet with the scaled time $t/T$ [Eq.~(\ref{eq:scales})] for different values of the scaled initial droplet radii $R_0/L_0$ and at a fixed Weber number $We = 30$. (b) Log-log plot of the scaled decay time $\tau_{\xi}$ versus $R_0/L_0$; the black line represents the power-law relationship $\tau_{\xi} \sim [R_0/L_0]^{3/2}$. (c) Log-log plots of the complementary CPDF of the first-passage time $\tau$ (see text). (d) Semi-log plots of the PDF of $\cos (\theta)$, with $\theta$ the angle between $\bm u$ and the normal to the droplet's interface (inset: a schematic diagram illustrating the angle $\theta$).}
\end{figure*}
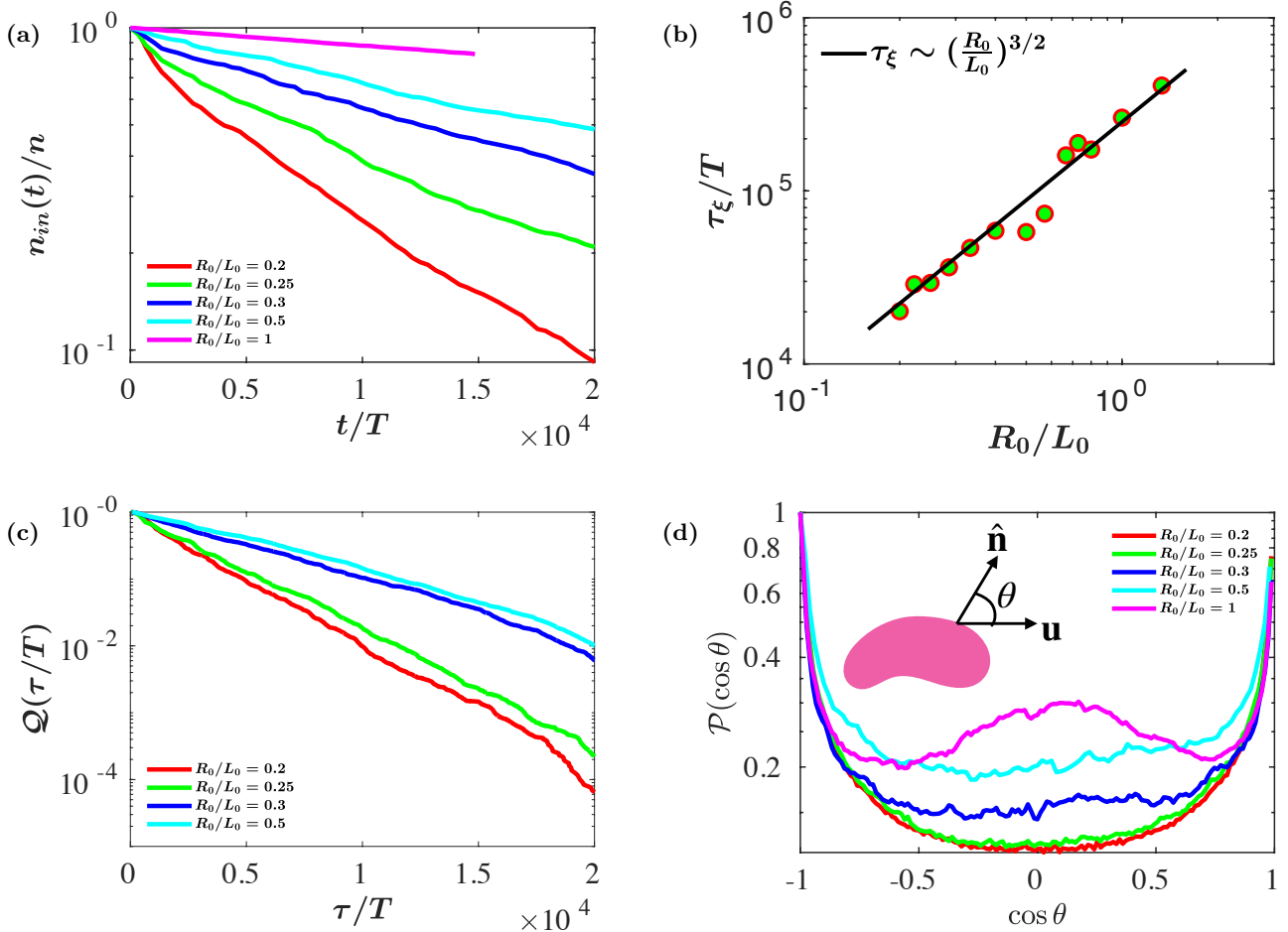
We depict the spatiotemporal evolution of tracers by using pseudocolor plots of vorticity fields overlaid with the $c = 0.5$ contour and tracer positions in Fig.~\ref{fig:phi_tracers}, at various representative values of the simulation time $t$. [See the Videos V1-V3 in Appendix~\ref{sec:videos} for the full spatiotemporal evolution.] For clarity, we limit the visualization to $6000$ particles, although we use $n = 1024 \times 128 = 131072$ tracer particles in our DNSs for statistical analysis. If the tracers are not confined within a droplet, as shown in Figs.\ref{fig:phi_tracers}(a)-(d), they disperse rapidly throughout the entire simulation domain. However, when we initiate our DNS with tracers inside a droplet, they remain entrapped within the droplet for a significant duration before emerging from it eventually, as we illustrate in Figs.\ref{fig:phi_tracers}(e)-(h) and (i)-(l) for droplets with radii $R_0/L_0 = 0.25$ and $R_0/L_0 = 0.5$, respectively. By comparing Figs.~\ref{fig:phi_tracers}(b), (f), and (j), captured at the same simulation time $t = 15$, we see how rapidly the tracers spread out without a droplet [Fig.~\ref{fig:phi_tracers}(b)] and how their spreading is constrained by the droplet boundary [Figs.~\ref{fig:phi_tracers}(f) and (j)], so they move along with the droplet. Clearly, the droplet interface acts as a transport barrier. \\

In the limit of an infinitely thin interface, with a high value of $\sigma$, no Lagrangian tracers should come out of the droplet. In the CHNS system we consider, the interface is diffuse; therefore, tracers do escape from droplets over a sufficiently long period of time. The narrower the interface, the longer is this time. We illustrate below how the mean droplet-escape time changes with the width of the CHNS interface; and we discuss the numerical subtleties of taking the limit of vanishing interface thickness.

To quantify the impact of the interfacial barrier on particle transport, we present semi-log plots, in Fig.~\ref{fig:n_tau}(a), of the fraction of particles inside the droplet $n_{in}(t)/n$ versus time $t$. For large $t$, these plots are consistent with the exponential decay
\begin{equation}
    \frac{n_{in}(t)}{n} \sim \exp(-t/\tau_{\xi})\,,
    \label{eq:decay}
\end{equation}
where the decay time $\tau_{\xi}$ depends on the scaled droplet radius $R_0/L_0$; we hold the Weber number at $We = 30$. In Fig.~\ref{fig:n_tau}(b) we give a log-log plot of $\tau_{\xi}$ versus $R_0/L_0$; this is consistent with [Fig.~\ref{fig:n_tau}(b)]
\begin{equation}
    \tau_{\xi} \sim \left[\frac{R_0}{L_0}\right]^{3/2}\,.
    \label{eq:tauxiR0}
\end{equation}
We obtain $\tau_{\xi}$ from a local-slope analysis of the exponentially decaying tails of the curves in Fig.~\ref{fig:n_tau}(a) [see Fig.~\ref{fig:slope} in Appendix~\ref{sec:slope} for details]. 

We have conducted two illustrative simulations with particles distributed uniformly at the centre of the simulation box within a radius $r/L_0 = 0.5$. (a) In the first simulation, we consider no droplet, effectively employing a hypothetical circular boundary (as depicted in Figs.\ref{fig:phi_tracers}(a)). (b) In the second simulation, we introduce a droplet into the system (as shown in Fig.\ref{fig:phi_tracers}(i)). At a simulation time of $t = 15$, the majority of the particles successfully cross the hypothetical boundary, with a radius of $r/L_0 = 0.5$, in the absence of a droplet, as illustrated in Fig.\ref{fig:phi_tracers}(b). In stark contrast, the presence of the droplet boundary in Fig.\ref{fig:phi_tracers}(j) acts as a transport barrier for the particles, causing them to be carried along with the droplet. 
\begin{figure*}
{
\includegraphics[width=\textwidth]{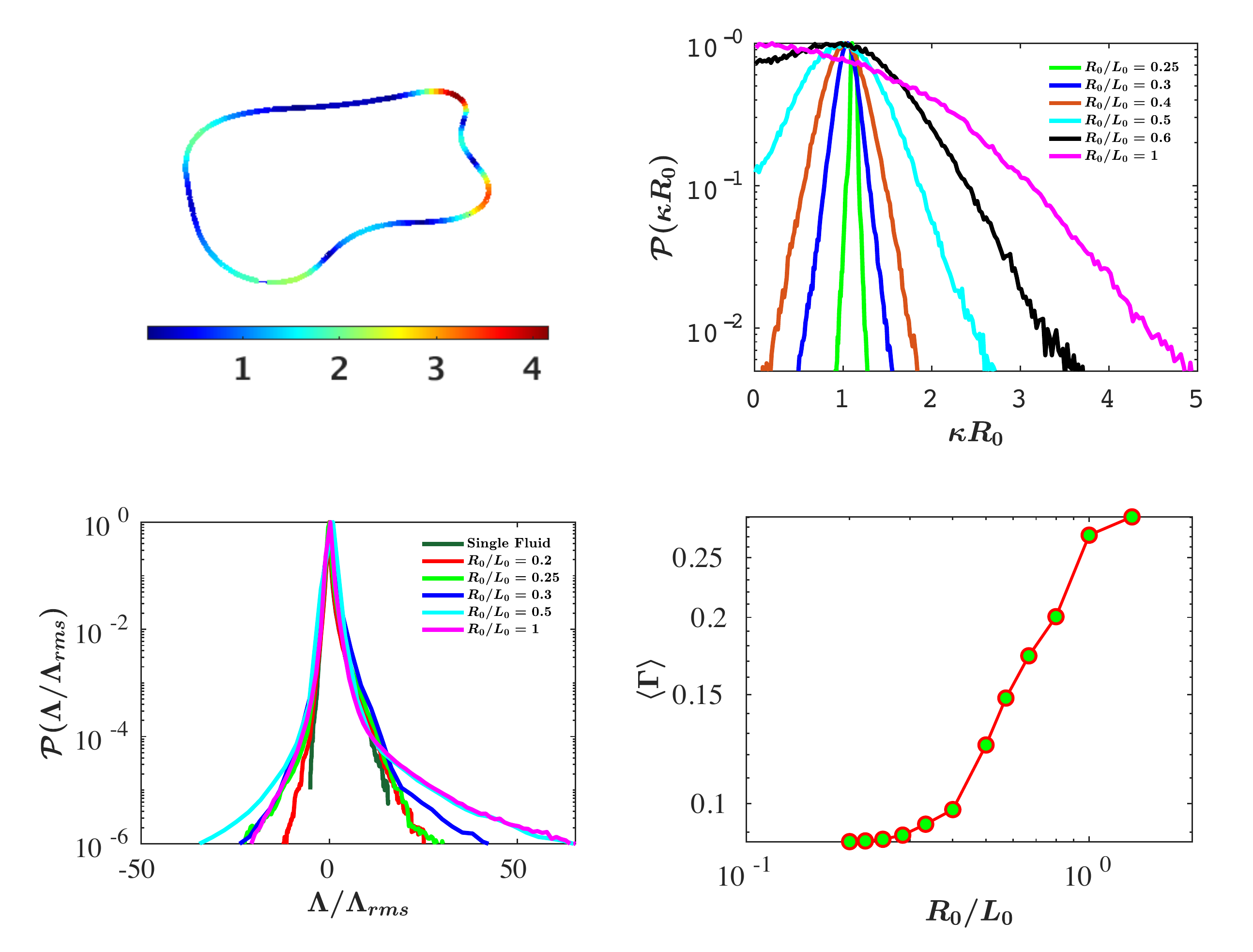}
\put(-495,350){\rm {\bf(a)}}
\put(-255,350){\rm {\bf(b)}}
\put(-495,160){\rm {\bf(c)}}
\put(-255,160){\rm {\bf(d)}}
}

\caption{\label{fig:kappa} (a) A pseudocolor plot of the normalized curvature $\kappa R_0$ along the droplet boundary (the $c = 0.5$ contour) at a representative time. (b) Semi-log plots of the PDF of $\kappa R_0$ for different values of $R_0/L_0$. (c) Semi-log plots of the PDF of the Okubo-Weiss parameter $\Lambda$ (see text) for different radii; these PDFs are normalized with the root-mean-square values of $\Lambda$. (d) Plot showing the time-averaged deformation $\langle\Gamma(t)\rangle_t$ (see text) of the droplet for different scaled radii $R_0/L_0$.}
\end{figure*}
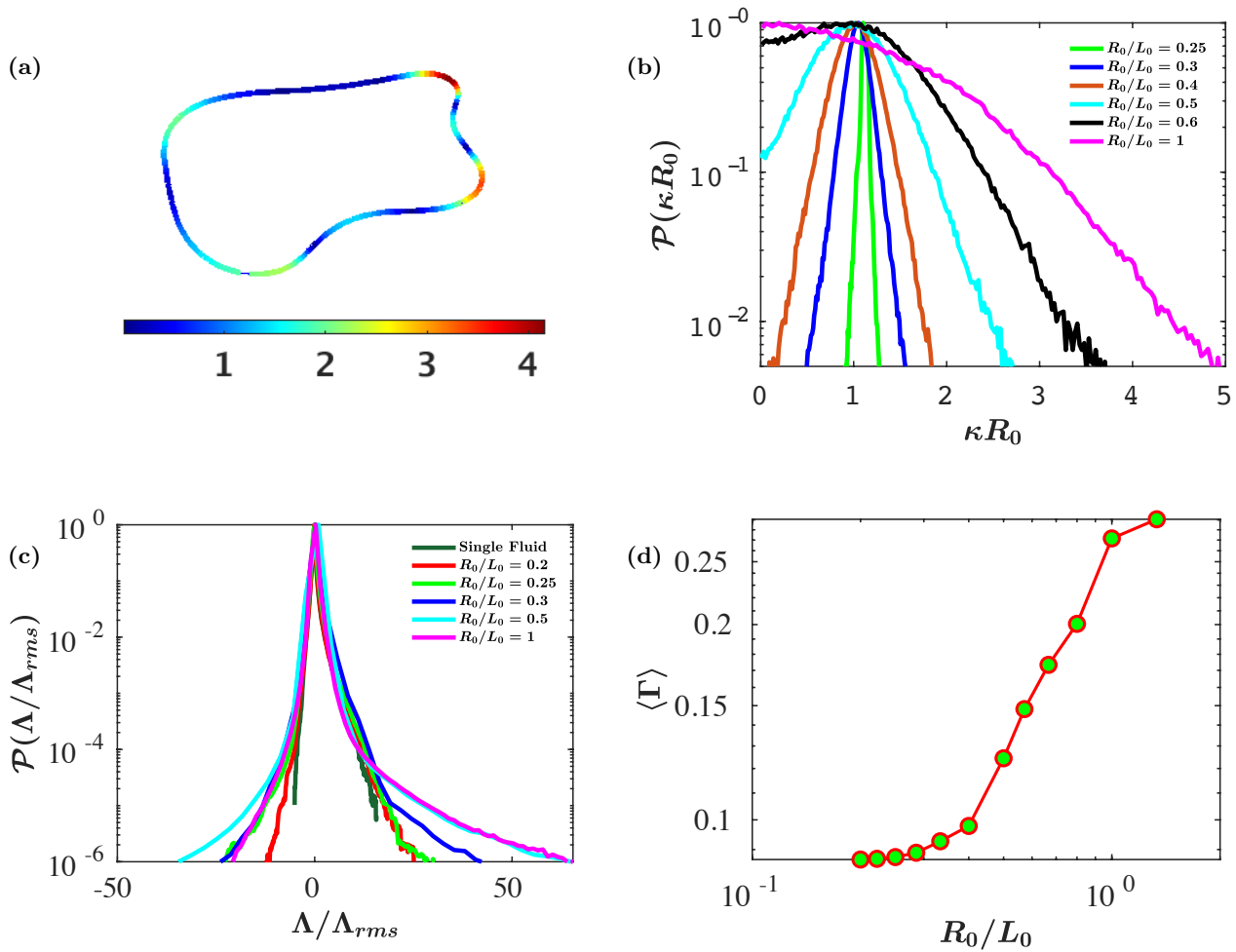
To characterise the time particles spend inside a droplet, or a hypothetical circular boundary, we calculate their first-passage-time $\tau$, which is the time required by a particle to cross the interface of the droplet or the boundary for the first time [cf. Ref.~\cite{verma2020first} for conventional fluid turbulence]. We determine the complementary cumulative probability distribution function (CPDF) of $\tau$, namely, $\mathcal Q(\tau)$, by using the rank-order method~\cite{verma2020first,mitra2005multiscaling,perlekar2011persistence},
for various droplet radii. In Fig.~\ref{fig:n_tau}(c) we give semi-log plots of $\mathcal Q(\tau)$ whose tails 
decay exponentially for all the radii we consider. We find that the average first-passage time $\langle \tau \rangle$ for the droplet with a scaled radius $R_0/L_0 = 0.5$ is approximately $300$ times longer than that for
the case of a hypothetical circular boundary with a scaled radius of $r/L_0 = 0.5$.

Let us now examine the PDF $\mathcal{P}(\cos \theta)$ of $\cos \theta$, where $\theta$ is the angle between the velocity vector and the normal to the interface, as illustrated in the schematic diagram of a droplet (see the inset of Fig.\ref{fig:n_tau}(d)):
\begin{equation}
    \cos (\theta) \equiv \frac{\hat n \cdot \bm u}{|\hat{n}||\bm u|}\,;\;\;
    \hat n = -\frac{\nabla c}{|\nabla c|}\,.
    \label{eq:costheta}
\end{equation}
 We plot $\mathcal{P}(\cos (\theta))$ in Fig.~\ref{fig:n_tau}(d) for various values of the scaled droplet radius $R_0/L_0$ of the droplet. We find that, for small values of $R_0/L_0$, the dominant peaks in $\mathcal{P}(\cos \theta)$ occur at $\cos (\theta) \simeq \pm 1$; this indicates that there is a tendency of velocity vectors to align or anti-align with $\hat n$. 
 By contrast, for large values of $R_0/L_0$, in addition the peaks at $\cos \theta \simeq \pm 1$, $\mathcal{P}(\cos \theta)$ exhibits a broad maximum around $\cos \theta \simeq 0$. Thus, there is an increase in the probability that velocity vectors are aligned tangentially to the interface and not only perpendicular to it, so the particles are less likely to cross the droplet interface if $R_0/L_0$ is large than when it is small. This suggests that, for droplets, it is useful to define a Cahn number $Cn_d \equiv \epsilon/R_0$, which is based on the droplet radius and decreases as $R_0$ increases. Thus,
 as $R_0$ increases, $Cn_d$ tends $0$, so $\tau_{\xi}$ increases (see the discussion below about the limit of vanishing interface thickness).  
 


\begin{figure*}
{
\includegraphics[width=\textwidth]{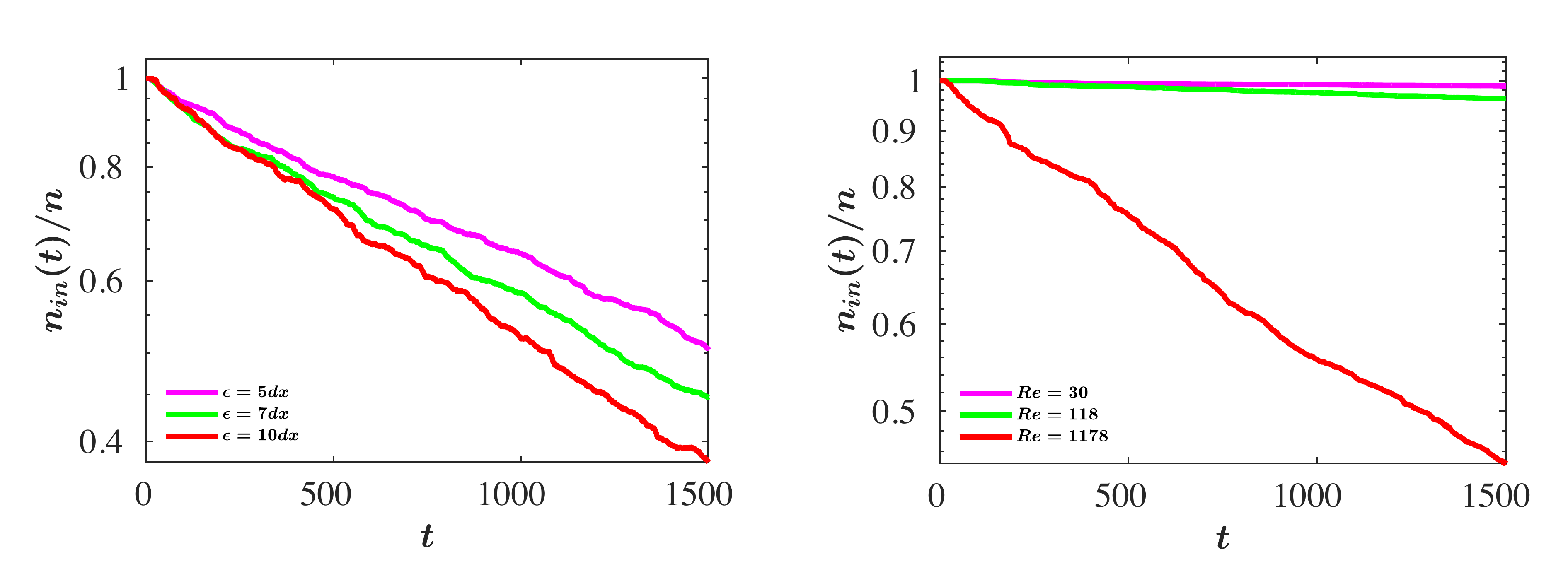}
\put(-500,160){\rm {\bf(a)}}
\put(-250,160){\rm {\bf(b)}}
}

\caption{\label{fig:eps_Re} Semi-log plots versus simulation time $t$ of the fraction $n_{in}(t)/n$ for (a) different values of interface widths $\epsilon$ (with 
fixed $Re = 1178$ and $R_0/L_0 = 0.5$) and (b) different values of the Reynolds number $Re$ (with fixed Cahn number $Cn=0.01$ and $R_0/L_0 = 0.5$).}
\end{figure*}
Next we characterise the interface, which acts as a transport barrier, by determining its scaled local curvature $\kappa R_0$. For this we use the local-curvature algorithm in MATLAB~\cite{MATLAB} to calculate the local curvature $\kappa$ along the droplet's interface, the $c = 0.5$ contour. In Fig.~\ref{fig:kappa}(a), we present an illustrative plot of $\kappa R_0$ along this contour for a droplet with a radius of $R_0/L_0 = 1$ at a representative time; the colors help us to distinguish regions of high and low curvature. To quantify curvature fluctuations, we calculate the PDFs of $\kappa R_0$ for different droplet scaled radii $R_0/L_0$ [semi-log plots in Fig.~\ref{fig:kappa}(b)]. These PDFs reveal that the larger the droplets the more significant their fluctuations, which appear to be linked to the effectiveness of large droplets as transport barriers. 

Curvature fluctuations are intrinsically linked to the deformation of the droplets. To quantify this deformation, we define a deformation parameter as follows~\cite{padhan2023activity,Pal_2016}:
\begin{eqnarray}
    \Gamma(t) = \frac{S(t)}{S_0(t)} - 1\,,
\end{eqnarray}
where $S(t)$ is the perimeter of the droplet at time $t$, and $S_0(t)$ denotes the perimeter of a circular droplet with an equivalent area at the same time. In Fig.~\ref{fig:kappa}(d), we present the mean deformation $\langle\Gamma(t)\rangle_t$ for various droplet radii; and we confirm that larger the droplet the more its deformation. 

Not only do large droplets deform more than small ones, but they also induce significant changes in the flow topology. To investigate these droplet-induced alterations in the flow topology, we calculate the Okubo-Weiss parameter $\Lambda$, which measures the relative importance of flow strain ($\Lambda > 0$) and vorticity ($\Lambda < 0$) and is defined as follows~\cite{weiss1991dynamics,okubo1970horizontal,soligo2020effect,pandit2017overview,shivamoggi2022okubo}:
\begin{eqnarray}
    \Lambda &=& D^2 - \Omega^2\,;\;\;
    D = \frac{\nabla \bm u + \nabla \bm u^T}{2}\,;\nonumber\\
    \Omega &=& \frac{\nabla \bm u - \nabla \bm u^T}{2}\,;
\end{eqnarray}
$D$ and $\Omega$ are the rate-of-deformation and rate-of-rotation tensor, respectively. In Fig.~\ref{fig:kappa}(c), we present semi-log plots of the PDFs of $\Lambda$ for various values of $R_0/L_0$. These PDFs 
reveal that the larger the droplets the more they affect $\Lambda$; in particular, these droplets enhance the tails of the PDFs of the Okubo-Weiss parameter. 

To illustrate how the mean droplet-escape time $\tau_\xi$ [see Eq.~(\ref{eq:decay})] changes with the width of the CHNS interface, we have carried out representative DNS runs with different values of the interfacial width $\epsilon$. The semi-log plots of the fraction $n_{in}(t)/n$ versus time $t$ in Fig.~\ref{fig:eps_Re} (a) show clearly that $\tau_\xi$ increases as $\epsilon$ decreases (with the Reynolds number held fixed at $Re = 1178$). A similar plot [Fig.~\ref{fig:eps_Re} (b)], with a fixed interface width characterised by the Cahn number $Cn=0.01$, shows that $\tau_\xi$ decreases as the 
Reynolds number $Re$ increases. We expect that $\tau_\xi \to \infty$, in the limit of an infinitely thin interface, i.e., $Cn \to 0$; however, it is numerically challenging to approach this limit in the CHNS system at large $Re$.

\section{Conclusions}
\label{sec:conclusions}

We have shown that the CHNS system provides a natural theoretical framework for examining how interfaces in turbulent binary-fluid flows act as transport barriers. Lagrangian tracer particles and their spreading in such flows play crucial roles in our quantitative characterisations of these transport barriers. In particular, the fraction of particles inside a droplet decays exponentially and the decay time $\tau_{\xi} \sim R_0^{2/3}$. 
The mean first-passage time $\langle \tau \rangle$, which we obtain from the CPDF of $\tau$, increases dramatically if the tracers are inside a droplet relative to its value without a droplet.  Perimeter and curvature fluctuations of a droplet's interface increase with $R_0/L_0$ and appear to enhance the ability of the interface to act as a transport barrier. Our studies also uncover the role of the PDF of $\cos(\theta)$ in enhancing the efficacy of this barrier. Our study of interfacial transport barriers in 2D binary-fluid flows can be generalised, \textit{mutatis mutandis}, to the three-dimensional (3D) CHNS system, to multi-fluid flows [see, e.g., Refs.~\cite{Jacqmin_1999} and \cite{padhan2023unveiling}], and to active-binary-fluid systems [see, e.g., Ref.~\cite{padhan2023activity}].  

Transport barriers occur in various turbulent flows. Important examples are found in systems that show staircases such as those that occur in plasma physics [see, e.g., Refs.~\cite{diamond1995dynamics} and \cite{ashourvan2017emergence}] and in double-diffusive convection [see, e.g., Refs.~\cite{radko2013double} and ~\cite{ouillon2020settling}]. Analogies between the 2D CHNS system and 2D MHD have been explored in detail in Refs.~\cite{fan2016cascades} and ~\cite{fan2018chns}. Therefore, transport barriers in systems with such staircases can be studied fruitfully by using the Lagrangian methods that we have developed in our CHNS-based study of interfaces as transport barriers in binary-fluid turbulence.

\section*{Acknowledgments}
We thank J.K. Alageshan, A.K. Verma, and K.V. Kiran for valuable discussions and the Science and Engineering Research Board (SERB) and the National Supercomputing Mission (NSM) [Grant No. $\text{DST/NSM/R\&D\_HPC\_Applications/2021/34}$] India for support, and the Supercomputer Education and Research Centre (IISc) for computational resources.  

\section*{Data and code availability}

Data from this study and the computer scripts can be obtained from the authors upon reasonable request.

\section*{Conflicts of Interest}
No conflicts of interests, financial or otherwise, are declared by the authors.

\section*{Author Contributions} NBP and RP planned the research and analysed the numerical data; NBP carried out the calculations and prepared the tables, figures, and the draft of the manuscript; NBP and RP then revised the manuscript in detail and approved the final version.

\appendix
\section{Local-slope analysis}
\label{sec:slope}

\begin{figure*}
{
\includegraphics[width=\textwidth]{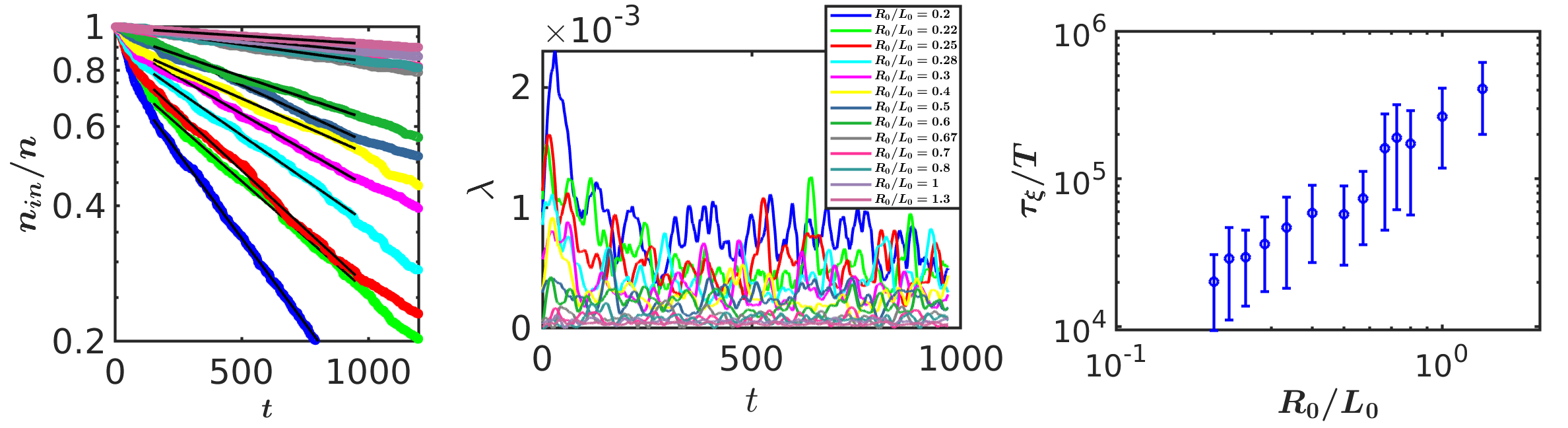}
\put(-510,125){\rm {\bf(a)}}
\put(-360,125){\rm {\bf(b)}}
\put(-185,125){\rm {\bf(c)}}
}

\caption{\label{fig:slope} (a) Semi-log plots of the fraction of particles inside the droplet $n_{in}(t)/n$ versus time $t$, illustrating the regions (black lines) within which we evaluate the local slopes for different scaled radii $R_0/L_0$. (b) Plots depicting the local slopes $\lambda=\tau_{\xi}/T$ over selected regions in (a). (c) Plot of the local slopes $\tau_{\xi}/T$ versus $R_0/L_0$, with one-standard-deviations error bars; the average value of the slope yields the decay time $\tau_{\xi}$ that we report in the main part of this paper.}
\end{figure*}

We obtain $\tau_{\xi}$ from a local-slope analysis of the exponentially decaying tails of the curves in Fig.~\ref{fig:n_tau}(a); we illustrate this analysis in Fig.~\ref{fig:slope}.

\section{Videos}
\label{sec:videos}
\begin{enumerate}
\item Video V1: Video showing the spatiotemporal evolution of the pseudocolor plots of the vorticity field overlaid with positions of the tracer particles (black points) as shown in  Fig.~\ref{fig:phi_tracers}(a)-(d) in the absence of droplets.
Video URL: \url{https://youtu.be/aXIuJAcYwac}
\item Video V2: Video showing the spatiotemporal evolution of the pseudocolor plots of the  vorticity field overlaid with the $c = 0.5$ contour (magenta line) and positions of the tracer particles as shown in  Fig.~\ref{fig:phi_tracers}(e)-(h) for the droplet of radius $R_0/L_0 = 0.25$. Video URL: \url{https://youtu.be/tAjNXb6Pz1I}
\item Video V3: Video showing the spatiotemporal evolution of the pseudocolor plots of the vorticity field overlaid with the $c = 0.5$ contour (magenta line) and positions of the tracer particles as shown in  Fig.~\ref{fig:phi_tracers}(i)-(l) for the droplet of radius $R_0/L_0 = 0.5$ . Video URL: \url{https://youtu.be/3wIBDXGkTvA}
\end{enumerate}

\nocite{*}

\bibliography{main}

\providecommand{\noopsort}[1]{}\providecommand{\singleletter}[1]{#1}%
\begin{thebibliography}{60}%
\makeatletter
\providecommand \@ifxundefined [1]{%
 \@ifx{#1\undefined}
}%
\providecommand \@ifnum [1]{%
 \ifnum #1\expandafter \@firstoftwo
 \else \expandafter \@secondoftwo
 \fi
}%
\providecommand \@ifx [1]{%
 \ifx #1\expandafter \@firstoftwo
 \else \expandafter \@secondoftwo
 \fi
}%
\providecommand \natexlab [1]{#1}%
\providecommand \enquote  [1]{``#1''}%
\providecommand \bibnamefont  [1]{#1}%
\providecommand \bibfnamefont [1]{#1}%
\providecommand \citenamefont [1]{#1}%
\providecommand \href@noop [0]{\@secondoftwo}%
\providecommand \href [0]{\begingroup \@sanitize@url \@href}%
\providecommand \@href[1]{\@@startlink{#1}\@@href}%
\providecommand \@@href[1]{\endgroup#1\@@endlink}%
\providecommand \@sanitize@url [0]{\catcode `\\12\catcode `\$12\catcode `\&12\catcode `\#12\catcode `\^12\catcode `\_12\catcode `\%12\relax}%
\providecommand \@@startlink[1]{}%
\providecommand \@@endlink[0]{}%
\providecommand \url  [0]{\begingroup\@sanitize@url \@url }%
\providecommand \@url [1]{\endgroup\@href {#1}{\urlprefix }}%
\providecommand \urlprefix  [0]{URL }%
\providecommand \Eprint [0]{\href }%
\providecommand \doibase [0]{https://doi.org/}%
\providecommand \selectlanguage [0]{\@gobble}%
\providecommand \bibinfo  [0]{\@secondoftwo}%
\providecommand \bibfield  [0]{\@secondoftwo}%
\providecommand \translation [1]{[#1]}%
\providecommand \BibitemOpen [0]{}%
\providecommand \bibitemStop [0]{}%
\providecommand \bibitemNoStop [0]{.\EOS\space}%
\providecommand \EOS [0]{\spacefactor3000\relax}%
\providecommand \BibitemShut  [1]{\csname bibitem#1\endcsname}%
\let\auto@bib@innerbib\@empty
\bibitem [{\citenamefont {Scardovelli}\ and\ \citenamefont {Zaleski}(1999)}]{scardovelli1999direct}%
  \BibitemOpen
  \bibfield  {author} {\bibinfo {author} {\bibfnamefont {R.}~\bibnamefont {Scardovelli}}\ and\ \bibinfo {author} {\bibfnamefont {S.}~\bibnamefont {Zaleski}},\ }\bibfield  {title} {\enquote {\bibinfo {title} {Direct numerical simulation of free-surface and interfacial flow},}\ }\href@noop {} {\bibfield  {journal} {\bibinfo  {journal} {Annual review of fluid mechanics}\ }\textbf {\bibinfo {volume} {31}},\ \bibinfo {pages} {567--603} (\bibinfo {year} {1999})}\BibitemShut {NoStop}%
\bibitem [{\citenamefont {Rider}\ and\ \citenamefont {Kothe}(1998)}]{rider1998reconstructing}%
  \BibitemOpen
  \bibfield  {author} {\bibinfo {author} {\bibfnamefont {W.~J.}\ \bibnamefont {Rider}}\ and\ \bibinfo {author} {\bibfnamefont {D.~B.}\ \bibnamefont {Kothe}},\ }\bibfield  {title} {\enquote {\bibinfo {title} {Reconstructing volume tracking},}\ }\href@noop {} {\bibfield  {journal} {\bibinfo  {journal} {Journal of computational physics}\ }\textbf {\bibinfo {volume} {141}},\ \bibinfo {pages} {112--152} (\bibinfo {year} {1998})}\BibitemShut {NoStop}%
\bibitem [{\citenamefont {Roccon}, \citenamefont {Zonta},\ and\ \citenamefont {Soldati}(2023)}]{roccon2023phase}%
  \BibitemOpen
  \bibfield  {author} {\bibinfo {author} {\bibfnamefont {A.}~\bibnamefont {Roccon}}, \bibinfo {author} {\bibfnamefont {F.}~\bibnamefont {Zonta}},\ and\ \bibinfo {author} {\bibfnamefont {A.}~\bibnamefont {Soldati}},\ }\bibfield  {title} {\enquote {\bibinfo {title} {Phase-field modeling of complex interface dynamics in drop-laden turbulence},}\ }\href@noop {} {\bibfield  {journal} {\bibinfo  {journal} {Physical Review Fluids}\ }\textbf {\bibinfo {volume} {8}},\ \bibinfo {pages} {090501} (\bibinfo {year} {2023})}\BibitemShut {NoStop}%
\bibitem [{\citenamefont {Dodd}\ and\ \citenamefont {Jofre}(2019)}]{dodd2019small}%
  \BibitemOpen
  \bibfield  {author} {\bibinfo {author} {\bibfnamefont {M.~S.}\ \bibnamefont {Dodd}}\ and\ \bibinfo {author} {\bibfnamefont {L.}~\bibnamefont {Jofre}},\ }\bibfield  {title} {\enquote {\bibinfo {title} {Small-scale flow topologies in decaying isotropic turbulence laden with finite-size droplets},}\ }\href@noop {} {\bibfield  {journal} {\bibinfo  {journal} {Physical Review Fluids}\ }\textbf {\bibinfo {volume} {4}},\ \bibinfo {pages} {064303} (\bibinfo {year} {2019})}\BibitemShut {NoStop}%
\bibitem [{\citenamefont {Soligo}, \citenamefont {Roccon},\ and\ \citenamefont {Soldati}(2020)}]{soligo2020effect}%
  \BibitemOpen
  \bibfield  {author} {\bibinfo {author} {\bibfnamefont {G.}~\bibnamefont {Soligo}}, \bibinfo {author} {\bibfnamefont {A.}~\bibnamefont {Roccon}},\ and\ \bibinfo {author} {\bibfnamefont {A.}~\bibnamefont {Soldati}},\ }\bibfield  {title} {\enquote {\bibinfo {title} {Effect of surfactant-laden droplets on turbulent flow topology},}\ }\href@noop {} {\bibfield  {journal} {\bibinfo  {journal} {Physical Review Fluids}\ }\textbf {\bibinfo {volume} {5}},\ \bibinfo {pages} {073606} (\bibinfo {year} {2020})}\BibitemShut {NoStop}%
\bibitem [{\citenamefont {Rosti}\ \emph {et~al.}(2019)\citenamefont {Rosti}, \citenamefont {Ge}, \citenamefont {Jain}, \citenamefont {Dodd},\ and\ \citenamefont {Brandt}}]{rosti2019droplets}%
  \BibitemOpen
  \bibfield  {author} {\bibinfo {author} {\bibfnamefont {M.~E.}\ \bibnamefont {Rosti}}, \bibinfo {author} {\bibfnamefont {Z.}~\bibnamefont {Ge}}, \bibinfo {author} {\bibfnamefont {S.~S.}\ \bibnamefont {Jain}}, \bibinfo {author} {\bibfnamefont {M.~S.}\ \bibnamefont {Dodd}},\ and\ \bibinfo {author} {\bibfnamefont {L.}~\bibnamefont {Brandt}},\ }\bibfield  {title} {\enquote {\bibinfo {title} {Droplets in homogeneous shear turbulence},}\ }\href@noop {} {\bibfield  {journal} {\bibinfo  {journal} {Journal of Fluid Mechanics}\ }\textbf {\bibinfo {volume} {876}},\ \bibinfo {pages} {962--984} (\bibinfo {year} {2019})}\BibitemShut {NoStop}%
\bibitem [{\citenamefont {Hajisharifi}, \citenamefont {Marchioli},\ and\ \citenamefont {Soldati}(2021)}]{hajisharifi2021particle}%
  \BibitemOpen
  \bibfield  {author} {\bibinfo {author} {\bibfnamefont {A.}~\bibnamefont {Hajisharifi}}, \bibinfo {author} {\bibfnamefont {C.}~\bibnamefont {Marchioli}},\ and\ \bibinfo {author} {\bibfnamefont {A.}~\bibnamefont {Soldati}},\ }\bibfield  {title} {\enquote {\bibinfo {title} {Particle capture by drops in turbulent flow},}\ }\href@noop {} {\bibfield  {journal} {\bibinfo  {journal} {Physical Review Fluids}\ }\textbf {\bibinfo {volume} {6}},\ \bibinfo {pages} {024303} (\bibinfo {year} {2021})}\BibitemShut {NoStop}%
\bibitem [{\citenamefont {Navier}(1823)}]{navier1823memoire}%
  \BibitemOpen
  \bibfield  {author} {\bibinfo {author} {\bibfnamefont {C.}~\bibnamefont {Navier}},\ }\bibfield  {title} {\enquote {\bibinfo {title} {M{\'e}moire sur les lois du mouvement des fluides},}\ }\href@noop {} {\bibfield  {journal} {\bibinfo  {journal} {M{\'e}moires de l’Acad{\'e}mie Royale des Sciences de l’Institut de France}\ }\textbf {\bibinfo {volume} {6}},\ \bibinfo {pages} {389--440} (\bibinfo {year} {1823})}\BibitemShut {NoStop}%
\bibitem [{\citenamefont {Doering}\ and\ \citenamefont {Gibbon}(1995)}]{doering1995applied}%
  \BibitemOpen
  \bibfield  {author} {\bibinfo {author} {\bibfnamefont {C.~R.}\ \bibnamefont {Doering}}\ and\ \bibinfo {author} {\bibfnamefont {J.~D.}\ \bibnamefont {Gibbon}},\ }\href@noop {} {\emph {\bibinfo {title} {Applied analysis of the Navier-Stokes equations}}},\ \bibinfo {number} {12}\ (\bibinfo  {publisher} {Cambridge university press},\ \bibinfo {year} {1995})\BibitemShut {NoStop}%
\bibitem [{\citenamefont {Cahn}\ and\ \citenamefont {Hilliard}(1958)}]{cahn1958free}%
  \BibitemOpen
  \bibfield  {author} {\bibinfo {author} {\bibfnamefont {J.~W.}\ \bibnamefont {Cahn}}\ and\ \bibinfo {author} {\bibfnamefont {J.~E.}\ \bibnamefont {Hilliard}},\ }\bibfield  {title} {\enquote {\bibinfo {title} {Free energy of a nonuniform system. i. interfacial free energy},}\ }\href@noop {} {\bibfield  {journal} {\bibinfo  {journal} {The Journal of chemical physics}\ }\textbf {\bibinfo {volume} {28}},\ \bibinfo {pages} {258--267} (\bibinfo {year} {1958})}\BibitemShut {NoStop}%
\bibitem [{\citenamefont {Cahn}\ and\ \citenamefont {Hilliard}(1959)}]{cahn1959free}%
  \BibitemOpen
  \bibfield  {author} {\bibinfo {author} {\bibfnamefont {J.~W.}\ \bibnamefont {Cahn}}\ and\ \bibinfo {author} {\bibfnamefont {J.~E.}\ \bibnamefont {Hilliard}},\ }\bibfield  {title} {\enquote {\bibinfo {title} {Free energy of a nonuniform system. iii. nucleation in a two-component incompressible fluid},}\ }\href@noop {} {\bibfield  {journal} {\bibinfo  {journal} {The Journal of chemical physics}\ }\textbf {\bibinfo {volume} {31}},\ \bibinfo {pages} {688--699} (\bibinfo {year} {1959})}\BibitemShut {NoStop}%
\bibitem [{\citenamefont {Cahn}(1961)}]{cahn1961spinodal}%
  \BibitemOpen
  \bibfield  {author} {\bibinfo {author} {\bibfnamefont {J.~W.}\ \bibnamefont {Cahn}},\ }\bibfield  {title} {\enquote {\bibinfo {title} {On spinodal decomposition},}\ }\href@noop {} {\bibfield  {journal} {\bibinfo  {journal} {Acta metallurgica}\ }\textbf {\bibinfo {volume} {9}},\ \bibinfo {pages} {795--801} (\bibinfo {year} {1961})}\BibitemShut {NoStop}%
\bibitem [{\citenamefont {Hohenberg}\ and\ \citenamefont {Halperin}(1977)}]{hohenberg1977theory}%
  \BibitemOpen
  \bibfield  {author} {\bibinfo {author} {\bibfnamefont {P.~C.}\ \bibnamefont {Hohenberg}}\ and\ \bibinfo {author} {\bibfnamefont {B.~I.}\ \bibnamefont {Halperin}},\ }\bibfield  {title} {\enquote {\bibinfo {title} {Theory of dynamic critical phenomena},}\ }\href@noop {} {\bibfield  {journal} {\bibinfo  {journal} {Reviews of Modern Physics}\ }\textbf {\bibinfo {volume} {49}},\ \bibinfo {pages} {435} (\bibinfo {year} {1977})}\BibitemShut {NoStop}%
\bibitem [{\citenamefont {Onuki}(2002)}]{onuki2002phase}%
  \BibitemOpen
  \bibfield  {author} {\bibinfo {author} {\bibfnamefont {A.}~\bibnamefont {Onuki}},\ }\href@noop {} {\emph {\bibinfo {title} {Phase transition dynamics}}}\ (\bibinfo  {publisher} {Cambridge University Press},\ \bibinfo {year} {2002})\BibitemShut {NoStop}%
\bibitem [{\citenamefont {Puri}\ and\ \citenamefont {Wadhawan}(2009)}]{puri2009kinetics}%
  \BibitemOpen
  \bibfield  {author} {\bibinfo {author} {\bibfnamefont {S.}~\bibnamefont {Puri}}\ and\ \bibinfo {author} {\bibfnamefont {V.}~\bibnamefont {Wadhawan}},\ }\href@noop {} {\emph {\bibinfo {title} {Kinetics of phase transitions}}}\ (\bibinfo  {publisher} {CRC press},\ \bibinfo {year} {2009})\BibitemShut {NoStop}%
\bibitem [{\citenamefont {Lee}\ \emph {et~al.}(2014)\citenamefont {Lee}, \citenamefont {Huh}, \citenamefont {Jeong}, \citenamefont {Shin}, \citenamefont {Yun},\ and\ \citenamefont {Kim}}]{lee2014physical}%
  \BibitemOpen
  \bibfield  {author} {\bibinfo {author} {\bibfnamefont {D.}~\bibnamefont {Lee}}, \bibinfo {author} {\bibfnamefont {J.-Y.}\ \bibnamefont {Huh}}, \bibinfo {author} {\bibfnamefont {D.}~\bibnamefont {Jeong}}, \bibinfo {author} {\bibfnamefont {J.}~\bibnamefont {Shin}}, \bibinfo {author} {\bibfnamefont {A.}~\bibnamefont {Yun}},\ and\ \bibinfo {author} {\bibfnamefont {J.}~\bibnamefont {Kim}},\ }\bibfield  {title} {\enquote {\bibinfo {title} {Physical, mathematical, and numerical derivations of the cahn--hilliard equation},}\ }\href@noop {} {\bibfield  {journal} {\bibinfo  {journal} {Computational Materials Science}\ }\textbf {\bibinfo {volume} {81}},\ \bibinfo {pages} {216--225} (\bibinfo {year} {2014})}\BibitemShut {NoStop}%
\bibitem [{\citenamefont {Jacqmin}(1999)}]{Jacqmin_1999}%
  \BibitemOpen
  \bibfield  {author} {\bibinfo {author} {\bibfnamefont {D.}~\bibnamefont {Jacqmin}},\ }\bibfield  {title} {\enquote {\bibinfo {title} {Calculation of two-phase navier--stokes flows using phase-field modeling},}\ }\href@noop {} {\bibfield  {journal} {\bibinfo  {journal} {Journal of computational physics}\ }\textbf {\bibinfo {volume} {155}},\ \bibinfo {pages} {96--127} (\bibinfo {year} {1999})}\BibitemShut {NoStop}%
\bibitem [{\citenamefont {Magaletti}\ \emph {et~al.}(2013)\citenamefont {Magaletti}, \citenamefont {Picano}, \citenamefont {Chinappi}, \citenamefont {Marino},\ and\ \citenamefont {Casciola}}]{Magaletti_2013}%
  \BibitemOpen
  \bibfield  {author} {\bibinfo {author} {\bibfnamefont {F.}~\bibnamefont {Magaletti}}, \bibinfo {author} {\bibfnamefont {F.}~\bibnamefont {Picano}}, \bibinfo {author} {\bibfnamefont {M.}~\bibnamefont {Chinappi}}, \bibinfo {author} {\bibfnamefont {L.}~\bibnamefont {Marino}},\ and\ \bibinfo {author} {\bibfnamefont {C.~M.}\ \bibnamefont {Casciola}},\ }\bibfield  {title} {\enquote {\bibinfo {title} {The sharp-interface limit of the cahn--hilliard/navier--stokes model for binary fluids},}\ }\href@noop {} {\bibfield  {journal} {\bibinfo  {journal} {Journal of Fluid Mechanics}\ }\textbf {\bibinfo {volume} {714}},\ \bibinfo {pages} {95--126} (\bibinfo {year} {2013})}\BibitemShut {NoStop}%
\bibitem [{\citenamefont {Boyer}\ and\ \citenamefont {Lapuerta}(2006)}]{Boyer_2006}%
  \BibitemOpen
  \bibfield  {author} {\bibinfo {author} {\bibfnamefont {F.}~\bibnamefont {Boyer}}\ and\ \bibinfo {author} {\bibfnamefont {C.}~\bibnamefont {Lapuerta}},\ }\bibfield  {title} {\enquote {\bibinfo {title} {Study of a three component cahn-hilliard flow model},}\ }\href@noop {} {\bibfield  {journal} {\bibinfo  {journal} {ESAIM: Mathematical Modelling and Numerical Analysis-Mod{\'e}lisation Math{\'e}matique et Analyse Num{\'e}rique}\ }\textbf {\bibinfo {volume} {40}},\ \bibinfo {pages} {653--687} (\bibinfo {year} {2006})}\BibitemShut {NoStop}%
\bibitem [{\citenamefont {Mirjalili}, \citenamefont {Jain},\ and\ \citenamefont {Mani}()}]{mirjalilimodeling}%
  \BibitemOpen
  \bibfield  {author} {\bibinfo {author} {\bibfnamefont {S.}~\bibnamefont {Mirjalili}}, \bibinfo {author} {\bibfnamefont {S.}~\bibnamefont {Jain}},\ and\ \bibinfo {author} {\bibfnamefont {A.}~\bibnamefont {Mani}},\ }\bibfield  {title} {\enquote {\bibinfo {title} {Modeling heat and mass transfer across interfaces in two-phase flows using phase-field methods},}\ }\href@noop {} {\ }\BibitemShut {NoStop}%
\bibitem [{\citenamefont {Celani}\ \emph {et~al.}(2009)\citenamefont {Celani}, \citenamefont {Mazzino}, \citenamefont {Muratore-Ginanneschi},\ and\ \citenamefont {Vozella}}]{celani2009phase}%
  \BibitemOpen
  \bibfield  {author} {\bibinfo {author} {\bibfnamefont {A.}~\bibnamefont {Celani}}, \bibinfo {author} {\bibfnamefont {A.}~\bibnamefont {Mazzino}}, \bibinfo {author} {\bibfnamefont {P.}~\bibnamefont {Muratore-Ginanneschi}},\ and\ \bibinfo {author} {\bibfnamefont {L.}~\bibnamefont {Vozella}},\ }\bibfield  {title} {\enquote {\bibinfo {title} {Phase-field model for the rayleigh--taylor instability of immiscible fluids},}\ }\href@noop {} {\bibfield  {journal} {\bibinfo  {journal} {Journal of Fluid Mechanics}\ }\textbf {\bibinfo {volume} {622}},\ \bibinfo {pages} {115--134} (\bibinfo {year} {2009})}\BibitemShut {NoStop}%
\bibitem [{\citenamefont {Lee}\ and\ \citenamefont {Kim}(2013)}]{lee2013numerical}%
  \BibitemOpen
  \bibfield  {author} {\bibinfo {author} {\bibfnamefont {H.~G.}\ \bibnamefont {Lee}}\ and\ \bibinfo {author} {\bibfnamefont {J.}~\bibnamefont {Kim}},\ }\bibfield  {title} {\enquote {\bibinfo {title} {Numerical simulation of the three-dimensional rayleigh--taylor instability},}\ }\href@noop {} {\bibfield  {journal} {\bibinfo  {journal} {Computers \& Mathematics with Applications}\ }\textbf {\bibinfo {volume} {66}},\ \bibinfo {pages} {1466--1474} (\bibinfo {year} {2013})}\BibitemShut {NoStop}%
\bibitem [{\citenamefont {Gibbon}\ \emph {et~al.}(2016)\citenamefont {Gibbon}, \citenamefont {Pal}, \citenamefont {Gupta},\ and\ \citenamefont {Pandit}}]{gibbon2016regularity}%
  \BibitemOpen
  \bibfield  {author} {\bibinfo {author} {\bibfnamefont {J.~D.}\ \bibnamefont {Gibbon}}, \bibinfo {author} {\bibfnamefont {N.}~\bibnamefont {Pal}}, \bibinfo {author} {\bibfnamefont {A.}~\bibnamefont {Gupta}},\ and\ \bibinfo {author} {\bibfnamefont {R.}~\bibnamefont {Pandit}},\ }\bibfield  {title} {\enquote {\bibinfo {title} {Regularity criterion for solutions of the three-dimensional cahn-hilliard-navier-stokes equations and associated computations},}\ }\href@noop {} {\bibfield  {journal} {\bibinfo  {journal} {Physical Review E}\ }\textbf {\bibinfo {volume} {94}},\ \bibinfo {pages} {063103} (\bibinfo {year} {2016})}\BibitemShut {NoStop}%
\bibitem [{\citenamefont {Lee}\ and\ \citenamefont {Kim}(2015)}]{lee2015two}%
  \BibitemOpen
  \bibfield  {author} {\bibinfo {author} {\bibfnamefont {H.~G.}\ \bibnamefont {Lee}}\ and\ \bibinfo {author} {\bibfnamefont {J.}~\bibnamefont {Kim}},\ }\bibfield  {title} {\enquote {\bibinfo {title} {Two-dimensional kelvin--helmholtz instabilities of multi-component fluids},}\ }\href@noop {} {\bibfield  {journal} {\bibinfo  {journal} {European Journal of Mechanics-B/Fluids}\ }\textbf {\bibinfo {volume} {49}},\ \bibinfo {pages} {77--88} (\bibinfo {year} {2015})}\BibitemShut {NoStop}%
\bibitem [{\citenamefont {Pal}\ \emph {et~al.}(2016)\citenamefont {Pal}, \citenamefont {Perlekar}, \citenamefont {Gupta},\ and\ \citenamefont {Pandit}}]{Pal_2016}%
  \BibitemOpen
  \bibfield  {author} {\bibinfo {author} {\bibfnamefont {N.}~\bibnamefont {Pal}}, \bibinfo {author} {\bibfnamefont {P.}~\bibnamefont {Perlekar}}, \bibinfo {author} {\bibfnamefont {A.}~\bibnamefont {Gupta}},\ and\ \bibinfo {author} {\bibfnamefont {R.}~\bibnamefont {Pandit}},\ }\bibfield  {title} {\enquote {\bibinfo {title} {Binary-fluid turbulence: Signatures of multifractal droplet dynamics and dissipation reduction},}\ }\href {https://doi.org/10.1103/PhysRevE.93.063115} {\bibfield  {journal} {\bibinfo  {journal} {Phys. Rev. E}\ }\textbf {\bibinfo {volume} {93}},\ \bibinfo {pages} {063115} (\bibinfo {year} {2016})}\BibitemShut {NoStop}%
\bibitem [{\citenamefont {Pal}(2016)}]{palphdthesis}%
  \BibitemOpen
  \bibfield  {author} {\bibinfo {author} {\bibfnamefont {N.}~\bibnamefont {Pal}},\ }\emph {\bibinfo {title} {Cahn-Hilliard-Navier-Stokes Investigations of Binary-Fluid Turbulence and Droplet Dynamics}},\ \href@noop {} {Ph.D. thesis},\ \bibinfo  {school} {Indian Institute of Science, Bangalore, India} (\bibinfo {year} {2016})\BibitemShut {NoStop}%
\bibitem [{\citenamefont {Padhan}\ and\ \citenamefont {Pandit}(2023{\natexlab{a}})}]{padhan2023activity}%
  \BibitemOpen
  \bibfield  {author} {\bibinfo {author} {\bibfnamefont {N.~B.}\ \bibnamefont {Padhan}}\ and\ \bibinfo {author} {\bibfnamefont {R.}~\bibnamefont {Pandit}},\ }\bibfield  {title} {\enquote {\bibinfo {title} {Activity-induced droplet propulsion and multifractality},}\ }\href@noop {} {\bibfield  {journal} {\bibinfo  {journal} {Physical Review Research}\ }\textbf {\bibinfo {volume} {5}},\ \bibinfo {pages} {L032013} (\bibinfo {year} {2023}{\natexlab{a}})}\BibitemShut {NoStop}%
\bibitem [{\citenamefont {Padhan}\ and\ \citenamefont {Pandit}(2023{\natexlab{b}})}]{padhan2023unveiling}%
  \BibitemOpen
  \bibfield  {author} {\bibinfo {author} {\bibfnamefont {N.~B.}\ \bibnamefont {Padhan}}\ and\ \bibinfo {author} {\bibfnamefont {R.}~\bibnamefont {Pandit}},\ }\bibfield  {title} {\enquote {\bibinfo {title} {Unveiling the spatiotemporal evolution of liquid-lens coalescence: Self-similarity, vortex quadrupoles, and turbulence in a three-phase fluid system},}\ }\href@noop {} {\bibfield  {journal} {\bibinfo  {journal} {arXiv preprint arXiv:2308.08993}\ } (\bibinfo {year} {2023}{\natexlab{b}})}\BibitemShut {NoStop}%
\bibitem [{\citenamefont {Yeung}(2002)}]{yeung2002lagrangian}%
  \BibitemOpen
  \bibfield  {author} {\bibinfo {author} {\bibfnamefont {P.}~\bibnamefont {Yeung}},\ }\bibfield  {title} {\enquote {\bibinfo {title} {Lagrangian investigations of turbulence},}\ }\href@noop {} {\bibfield  {journal} {\bibinfo  {journal} {Annual review of fluid mechanics}\ }\textbf {\bibinfo {volume} {34}},\ \bibinfo {pages} {115--142} (\bibinfo {year} {2002})}\BibitemShut {NoStop}%
\bibitem [{\citenamefont {Sreenivasan}\ and\ \citenamefont {Schumacher}(2010)}]{sreenivasan2010lagrangian}%
  \BibitemOpen
  \bibfield  {author} {\bibinfo {author} {\bibfnamefont {K.~R.}\ \bibnamefont {Sreenivasan}}\ and\ \bibinfo {author} {\bibfnamefont {J.}~\bibnamefont {Schumacher}},\ }\bibfield  {title} {\enquote {\bibinfo {title} {Lagrangian views on turbulent mixing of passive scalars},}\ }\href@noop {} {\bibfield  {journal} {\bibinfo  {journal} {Philosophical Transactions of the Royal Society A: Mathematical, Physical and Engineering Sciences}\ }\textbf {\bibinfo {volume} {368}},\ \bibinfo {pages} {1561--1577} (\bibinfo {year} {2010})}\BibitemShut {NoStop}%
\bibitem [{\citenamefont {Biferale}\ \emph {et~al.}(2006)\citenamefont {Biferale}, \citenamefont {Boffetta}, \citenamefont {Celani}, \citenamefont {Lanotte},\ and\ \citenamefont {Toschi}}]{biferale2006lagrangian}%
  \BibitemOpen
  \bibfield  {author} {\bibinfo {author} {\bibfnamefont {L.}~\bibnamefont {Biferale}}, \bibinfo {author} {\bibfnamefont {G.}~\bibnamefont {Boffetta}}, \bibinfo {author} {\bibfnamefont {A.}~\bibnamefont {Celani}}, \bibinfo {author} {\bibfnamefont {A.}~\bibnamefont {Lanotte}},\ and\ \bibinfo {author} {\bibfnamefont {F.}~\bibnamefont {Toschi}},\ }\bibfield  {title} {\enquote {\bibinfo {title} {Lagrangian statistics in fully developed turbulence},}\ }\href@noop {} {\bibfield  {journal} {\bibinfo  {journal} {Journal of Turbulence}\ ,\ \bibinfo {pages} {N6}} (\bibinfo {year} {2006})}\BibitemShut {NoStop}%
\bibitem [{\citenamefont {Homann}\ \emph {et~al.}(2007)\citenamefont {Homann}, \citenamefont {Grauer}, \citenamefont {Busse},\ and\ \citenamefont {M{\"u}ller}}]{homann2007lagrangian}%
  \BibitemOpen
  \bibfield  {author} {\bibinfo {author} {\bibfnamefont {H.}~\bibnamefont {Homann}}, \bibinfo {author} {\bibfnamefont {R.}~\bibnamefont {Grauer}}, \bibinfo {author} {\bibfnamefont {A.}~\bibnamefont {Busse}},\ and\ \bibinfo {author} {\bibfnamefont {W.-C.}\ \bibnamefont {M{\"u}ller}},\ }\bibfield  {title} {\enquote {\bibinfo {title} {Lagrangian statistics of navier--stokes and mhd turbulence},}\ }\href@noop {} {\bibfield  {journal} {\bibinfo  {journal} {Journal of Plasma Physics}\ }\textbf {\bibinfo {volume} {73}},\ \bibinfo {pages} {821--830} (\bibinfo {year} {2007})}\BibitemShut {NoStop}%
\bibitem [{\citenamefont {Toschi}\ and\ \citenamefont {Bodenschatz}(2009)}]{toschi2009lagrangian}%
  \BibitemOpen
  \bibfield  {author} {\bibinfo {author} {\bibfnamefont {F.}~\bibnamefont {Toschi}}\ and\ \bibinfo {author} {\bibfnamefont {E.}~\bibnamefont {Bodenschatz}},\ }\bibfield  {title} {\enquote {\bibinfo {title} {Lagrangian properties of particles in turbulence},}\ }\href@noop {} {\bibfield  {journal} {\bibinfo  {journal} {Annual review of fluid mechanics}\ }\textbf {\bibinfo {volume} {41}},\ \bibinfo {pages} {375--404} (\bibinfo {year} {2009})}\BibitemShut {NoStop}%
\bibitem [{\citenamefont {Pandit}, \citenamefont {Perlekar},\ and\ \citenamefont {Ray}(2009)}]{pandit2009statistical}%
  \BibitemOpen
  \bibfield  {author} {\bibinfo {author} {\bibfnamefont {R.}~\bibnamefont {Pandit}}, \bibinfo {author} {\bibfnamefont {P.}~\bibnamefont {Perlekar}},\ and\ \bibinfo {author} {\bibfnamefont {S.~S.}\ \bibnamefont {Ray}},\ }\bibfield  {title} {\enquote {\bibinfo {title} {Statistical properties of turbulence: an overview},}\ }\href@noop {} {\bibfield  {journal} {\bibinfo  {journal} {Pramana}\ }\textbf {\bibinfo {volume} {73}},\ \bibinfo {pages} {157--191} (\bibinfo {year} {2009})}\BibitemShut {NoStop}%
\bibitem [{\citenamefont {La~Mantia}\ \emph {et~al.}(2013)\citenamefont {La~Mantia}, \citenamefont {Duda}, \citenamefont {Rotter},\ and\ \citenamefont {Skrbek}}]{la2013lagrangian}%
  \BibitemOpen
  \bibfield  {author} {\bibinfo {author} {\bibfnamefont {M.}~\bibnamefont {La~Mantia}}, \bibinfo {author} {\bibfnamefont {D.}~\bibnamefont {Duda}}, \bibinfo {author} {\bibfnamefont {M.}~\bibnamefont {Rotter}},\ and\ \bibinfo {author} {\bibfnamefont {L.}~\bibnamefont {Skrbek}},\ }\bibfield  {title} {\enquote {\bibinfo {title} {Lagrangian accelerations of particles in superfluid turbulence},}\ }\href@noop {} {\bibfield  {journal} {\bibinfo  {journal} {Journal of Fluid Mechanics}\ }\textbf {\bibinfo {volume} {717}},\ \bibinfo {pages} {R9} (\bibinfo {year} {2013})}\BibitemShut {NoStop}%
\bibitem [{\citenamefont {Singh}, \citenamefont {Mukherjee},\ and\ \citenamefont {Ray}(2022)}]{singh2022lagrangian}%
  \BibitemOpen
  \bibfield  {author} {\bibinfo {author} {\bibfnamefont {R.~K.}\ \bibnamefont {Singh}}, \bibinfo {author} {\bibfnamefont {S.}~\bibnamefont {Mukherjee}},\ and\ \bibinfo {author} {\bibfnamefont {S.~S.}\ \bibnamefont {Ray}},\ }\bibfield  {title} {\enquote {\bibinfo {title} {Lagrangian manifestation of anomalies in active turbulence},}\ }\href@noop {} {\bibfield  {journal} {\bibinfo  {journal} {Physical Review Fluids}\ }\textbf {\bibinfo {volume} {7}},\ \bibinfo {pages} {033101} (\bibinfo {year} {2022})}\BibitemShut {NoStop}%
\bibitem [{\citenamefont {Kiran}\ \emph {et~al.}(2023)\citenamefont {Kiran}, \citenamefont {Gupta}, \citenamefont {Verma},\ and\ \citenamefont {Pandit}}]{kiran2023irreversibility}%
  \BibitemOpen
  \bibfield  {author} {\bibinfo {author} {\bibfnamefont {K.~V.}\ \bibnamefont {Kiran}}, \bibinfo {author} {\bibfnamefont {A.}~\bibnamefont {Gupta}}, \bibinfo {author} {\bibfnamefont {A.~K.}\ \bibnamefont {Verma}},\ and\ \bibinfo {author} {\bibfnamefont {R.}~\bibnamefont {Pandit}},\ }\bibfield  {title} {\enquote {\bibinfo {title} {Irreversibility in bacterial turbulence: Insights from the mean-bacterial-velocity model},}\ }\href@noop {} {\bibfield  {journal} {\bibinfo  {journal} {Physical Review Fluids}\ }\textbf {\bibinfo {volume} {8}},\ \bibinfo {pages} {023102} (\bibinfo {year} {2023})}\BibitemShut {NoStop}%
\bibitem [{\citenamefont {Fan}\ \emph {et~al.}(2016)\citenamefont {Fan}, \citenamefont {Diamond}, \citenamefont {Chac{\'o}n},\ and\ \citenamefont {Li}}]{fan2016cascades}%
  \BibitemOpen
  \bibfield  {author} {\bibinfo {author} {\bibfnamefont {X.}~\bibnamefont {Fan}}, \bibinfo {author} {\bibfnamefont {P.}~\bibnamefont {Diamond}}, \bibinfo {author} {\bibfnamefont {L.}~\bibnamefont {Chac{\'o}n}},\ and\ \bibinfo {author} {\bibfnamefont {H.}~\bibnamefont {Li}},\ }\bibfield  {title} {\enquote {\bibinfo {title} {Cascades and spectra of a turbulent spinodal decomposition in two-dimensional symmetric binary liquid mixtures},}\ }\href@noop {} {\bibfield  {journal} {\bibinfo  {journal} {Physical Review Fluids}\ }\textbf {\bibinfo {volume} {1}},\ \bibinfo {pages} {054403} (\bibinfo {year} {2016})}\BibitemShut {NoStop}%
\bibitem [{\citenamefont {Fan}, \citenamefont {Diamond},\ and\ \citenamefont {Chac{\'o}n}(2018)}]{fan2018chns}%
  \BibitemOpen
  \bibfield  {author} {\bibinfo {author} {\bibfnamefont {X.}~\bibnamefont {Fan}}, \bibinfo {author} {\bibfnamefont {P.}~\bibnamefont {Diamond}},\ and\ \bibinfo {author} {\bibfnamefont {L.}~\bibnamefont {Chac{\'o}n}},\ }\bibfield  {title} {\enquote {\bibinfo {title} {Chns: A case study of turbulence in elastic media},}\ }\href@noop {} {\bibfield  {journal} {\bibinfo  {journal} {Physics of Plasmas}\ }\textbf {\bibinfo {volume} {25}} (\bibinfo {year} {2018})}\BibitemShut {NoStop}%
\bibitem [{\citenamefont {Chaikin}, \citenamefont {Lubensky},\ and\ \citenamefont {Witten}(1995)}]{chaikin1995principles}%
  \BibitemOpen
  \bibfield  {author} {\bibinfo {author} {\bibfnamefont {P.~M.}\ \bibnamefont {Chaikin}}, \bibinfo {author} {\bibfnamefont {T.~C.}\ \bibnamefont {Lubensky}},\ and\ \bibinfo {author} {\bibfnamefont {T.~A.}\ \bibnamefont {Witten}},\ }\href@noop {} {\emph {\bibinfo {title} {Principles of condensed matter physics}}},\ Vol.~\bibinfo {volume} {10}\ (\bibinfo  {publisher} {Cambridge university press Cambridge},\ \bibinfo {year} {1995})\BibitemShut {NoStop}%
\bibitem [{\citenamefont {Jain}\ and\ \citenamefont {Mani}(2023)}]{jain2023computational}%
  \BibitemOpen
  \bibfield  {author} {\bibinfo {author} {\bibfnamefont {S.~S.}\ \bibnamefont {Jain}}\ and\ \bibinfo {author} {\bibfnamefont {A.}~\bibnamefont {Mani}},\ }\bibfield  {title} {\enquote {\bibinfo {title} {A computational model for transport of immiscible scalars in two-phase flows},}\ }\href@noop {} {\bibfield  {journal} {\bibinfo  {journal} {Journal of Computational Physics}\ }\textbf {\bibinfo {volume} {476}},\ \bibinfo {pages} {111843} (\bibinfo {year} {2023})}\BibitemShut {NoStop}%
\bibitem [{\citenamefont {Pandit}\ \emph {et~al.}(2017)\citenamefont {Pandit}, \citenamefont {Banerjee}, \citenamefont {Bhatnagar}, \citenamefont {Brachet}, \citenamefont {Gupta}, \citenamefont {Mitra}, \citenamefont {Pal}, \citenamefont {Perlekar}, \citenamefont {Ray}, \citenamefont {Shukla} \emph {et~al.}}]{pandit2017overview}%
  \BibitemOpen
  \bibfield  {author} {\bibinfo {author} {\bibfnamefont {R.}~\bibnamefont {Pandit}}, \bibinfo {author} {\bibfnamefont {D.}~\bibnamefont {Banerjee}}, \bibinfo {author} {\bibfnamefont {A.}~\bibnamefont {Bhatnagar}}, \bibinfo {author} {\bibfnamefont {M.}~\bibnamefont {Brachet}}, \bibinfo {author} {\bibfnamefont {A.}~\bibnamefont {Gupta}}, \bibinfo {author} {\bibfnamefont {D.}~\bibnamefont {Mitra}}, \bibinfo {author} {\bibfnamefont {N.}~\bibnamefont {Pal}}, \bibinfo {author} {\bibfnamefont {P.}~\bibnamefont {Perlekar}}, \bibinfo {author} {\bibfnamefont {S.~S.}\ \bibnamefont {Ray}}, \bibinfo {author} {\bibfnamefont {V.}~\bibnamefont {Shukla}}, \emph {et~al.},\ }\bibfield  {title} {\enquote {\bibinfo {title} {An overview of the statistical properties of two-dimensional turbulence in fluids with particles, conducting fluids, fluids with polymer additives, binary-fluid mixtures, and superfluids},}\ }\href@noop {} {\bibfield  {journal} {\bibinfo  {journal} {Physics of fluids}\ }\textbf {\bibinfo {volume} {29}}
  (\bibinfo {year} {2017})}\BibitemShut {NoStop}%
\bibitem [{\citenamefont {Canuto}\ \emph {et~al.}(2012)\citenamefont {Canuto}, \citenamefont {Hussaini}, \citenamefont {Quarteroni}, \citenamefont {Thomas~Jr} \emph {et~al.}}]{canuto2012spectral}%
  \BibitemOpen
  \bibfield  {author} {\bibinfo {author} {\bibfnamefont {C.}~\bibnamefont {Canuto}}, \bibinfo {author} {\bibfnamefont {M.~Y.}\ \bibnamefont {Hussaini}}, \bibinfo {author} {\bibfnamefont {A.}~\bibnamefont {Quarteroni}}, \bibinfo {author} {\bibfnamefont {A.}~\bibnamefont {Thomas~Jr}}, \emph {et~al.},\ }\href@noop {} {\emph {\bibinfo {title} {Spectral methods in fluid dynamics}}}\ (\bibinfo  {publisher} {Springer Science Business Media},\ \bibinfo {year} {2012})\BibitemShut {NoStop}%
\bibitem [{\citenamefont {Cox}\ and\ \citenamefont {Matthews}(2002)}]{cox2002exponential}%
  \BibitemOpen
  \bibfield  {author} {\bibinfo {author} {\bibfnamefont {S.~M.}\ \bibnamefont {Cox}}\ and\ \bibinfo {author} {\bibfnamefont {P.~C.}\ \bibnamefont {Matthews}},\ }\bibfield  {title} {\enquote {\bibinfo {title} {Exponential time differencing for stiff systems},}\ }\href@noop {} {\bibfield  {journal} {\bibinfo  {journal} {Journal of Computational Physics}\ }\textbf {\bibinfo {volume} {176}},\ \bibinfo {pages} {430--455} (\bibinfo {year} {2002})}\BibitemShut {NoStop}%
\bibitem [{\citenamefont {Lalescu}\ and\ \citenamefont {Wilczek}(2018)}]{lalescu2018tracer}%
  \BibitemOpen
  \bibfield  {author} {\bibinfo {author} {\bibfnamefont {C.~C.}\ \bibnamefont {Lalescu}}\ and\ \bibinfo {author} {\bibfnamefont {M.}~\bibnamefont {Wilczek}},\ }\bibfield  {title} {\enquote {\bibinfo {title} {How tracer particles sample the complexity of turbulence},}\ }\href@noop {} {\bibfield  {journal} {\bibinfo  {journal} {New Journal of Physics}\ }\textbf {\bibinfo {volume} {20}},\ \bibinfo {pages} {013001} (\bibinfo {year} {2018})}\BibitemShut {NoStop}%
\bibitem [{\citenamefont {Verma}\ \emph {et~al.}(2020)\citenamefont {Verma}, \citenamefont {Bhatnagar}, \citenamefont {Mitra},\ and\ \citenamefont {Pandit}}]{verma2020first}%
  \BibitemOpen
  \bibfield  {author} {\bibinfo {author} {\bibfnamefont {A.~K.}\ \bibnamefont {Verma}}, \bibinfo {author} {\bibfnamefont {A.}~\bibnamefont {Bhatnagar}}, \bibinfo {author} {\bibfnamefont {D.}~\bibnamefont {Mitra}},\ and\ \bibinfo {author} {\bibfnamefont {R.}~\bibnamefont {Pandit}},\ }\bibfield  {title} {\enquote {\bibinfo {title} {First-passage-time problem for tracers in turbulent flows applied to virus spreading},}\ }\href@noop {} {\bibfield  {journal} {\bibinfo  {journal} {Physical Review Research}\ }\textbf {\bibinfo {volume} {2}},\ \bibinfo {pages} {033239} (\bibinfo {year} {2020})}\BibitemShut {NoStop}%
\bibitem [{\citenamefont {Benzi}\ \emph {et~al.}(2010)\citenamefont {Benzi}, \citenamefont {Biferale}, \citenamefont {Fisher}, \citenamefont {Lamb},\ and\ \citenamefont {Toschi}}]{benzi2010inertial}%
  \BibitemOpen
  \bibfield  {author} {\bibinfo {author} {\bibfnamefont {R.}~\bibnamefont {Benzi}}, \bibinfo {author} {\bibfnamefont {L.}~\bibnamefont {Biferale}}, \bibinfo {author} {\bibfnamefont {R.}~\bibnamefont {Fisher}}, \bibinfo {author} {\bibfnamefont {D.}~\bibnamefont {Lamb}},\ and\ \bibinfo {author} {\bibfnamefont {F.}~\bibnamefont {Toschi}},\ }\bibfield  {title} {\enquote {\bibinfo {title} {Inertial range eulerian and lagrangian statistics from numerical simulations of isotropic turbulence},}\ }\href@noop {} {\bibfield  {journal} {\bibinfo  {journal} {Journal of Fluid Mechanics}\ }\textbf {\bibinfo {volume} {653}},\ \bibinfo {pages} {221--244} (\bibinfo {year} {2010})}\BibitemShut {NoStop}%
\bibitem [{\citenamefont {Mitra}\ \emph {et~al.}(2005)\citenamefont {Mitra}, \citenamefont {Bec}, \citenamefont {Pandit},\ and\ \citenamefont {Frisch}}]{mitra2005multiscaling}%
  \BibitemOpen
  \bibfield  {author} {\bibinfo {author} {\bibfnamefont {D.}~\bibnamefont {Mitra}}, \bibinfo {author} {\bibfnamefont {J.}~\bibnamefont {Bec}}, \bibinfo {author} {\bibfnamefont {R.}~\bibnamefont {Pandit}},\ and\ \bibinfo {author} {\bibfnamefont {U.}~\bibnamefont {Frisch}},\ }\bibfield  {title} {\enquote {\bibinfo {title} {Is multiscaling an artifact in the stochastically forced burgers equation?}}\ }\href@noop {} {\bibfield  {journal} {\bibinfo  {journal} {Physical review letters}\ }\textbf {\bibinfo {volume} {94}},\ \bibinfo {pages} {194501} (\bibinfo {year} {2005})}\BibitemShut {NoStop}%
\bibitem [{\citenamefont {Perlekar}\ \emph {et~al.}(2011)\citenamefont {Perlekar}, \citenamefont {Ray}, \citenamefont {Mitra},\ and\ \citenamefont {Pandit}}]{perlekar2011persistence}%
  \BibitemOpen
  \bibfield  {author} {\bibinfo {author} {\bibfnamefont {P.}~\bibnamefont {Perlekar}}, \bibinfo {author} {\bibfnamefont {S.~S.}\ \bibnamefont {Ray}}, \bibinfo {author} {\bibfnamefont {D.}~\bibnamefont {Mitra}},\ and\ \bibinfo {author} {\bibfnamefont {R.}~\bibnamefont {Pandit}},\ }\bibfield  {title} {\enquote {\bibinfo {title} {Persistence problem in two-dimensional fluid turbulence},}\ }\href@noop {} {\bibfield  {journal} {\bibinfo  {journal} {Physical review letters}\ }\textbf {\bibinfo {volume} {106}},\ \bibinfo {pages} {054501} (\bibinfo {year} {2011})}\BibitemShut {NoStop}%
\bibitem [{MAT(2023)}]{MATLAB}%
  \BibitemOpen
  \href@noop {} {\enquote {\bibinfo {title} {Curvature analysis},}\ }\bibinfo {howpublished} {\url{https://in.mathworks.com/matlabcentral/fileexchange/93175-interface-curvature?s_tid=ta_fx_results}} (\bibinfo {year} {2023})\BibitemShut {NoStop}%
\bibitem [{\citenamefont {Weiss}(1991)}]{weiss1991dynamics}%
  \BibitemOpen
  \bibfield  {author} {\bibinfo {author} {\bibfnamefont {J.}~\bibnamefont {Weiss}},\ }\bibfield  {title} {\enquote {\bibinfo {title} {The dynamics of enstrophy transfer in two-dimensional hydrodynamics},}\ }\href@noop {} {\bibfield  {journal} {\bibinfo  {journal} {Physica D: Nonlinear Phenomena}\ }\textbf {\bibinfo {volume} {48}},\ \bibinfo {pages} {273--294} (\bibinfo {year} {1991})}\BibitemShut {NoStop}%
\bibitem [{\citenamefont {Okubo}(1970)}]{okubo1970horizontal}%
  \BibitemOpen
  \bibfield  {author} {\bibinfo {author} {\bibfnamefont {A.}~\bibnamefont {Okubo}},\ }\bibfield  {title} {\enquote {\bibinfo {title} {Horizontal dispersion of floatable particles in the vicinity of velocity singularities such as convergences},}\ }in\ \href@noop {} {\emph {\bibinfo {booktitle} {Deep sea research and oceanographic abstracts}}},\ Vol.~\bibinfo {volume} {17}\ (\bibinfo {organization} {Elsevier},\ \bibinfo {year} {1970})\ pp.\ \bibinfo {pages} {445--454}\BibitemShut {NoStop}%
\bibitem [{\citenamefont {Shivamoggi}, \citenamefont {van Heijst},\ and\ \citenamefont {Kamp}(2022)}]{shivamoggi2022okubo}%
  \BibitemOpen
  \bibfield  {author} {\bibinfo {author} {\bibfnamefont {B.}~\bibnamefont {Shivamoggi}}, \bibinfo {author} {\bibfnamefont {G.}~\bibnamefont {van Heijst}},\ and\ \bibinfo {author} {\bibfnamefont {L.}~\bibnamefont {Kamp}},\ }\bibfield  {title} {\enquote {\bibinfo {title} {The okubo--weiss criterion in hydrodynamic flows: geometric aspects and further extension},}\ }\href@noop {} {\bibfield  {journal} {\bibinfo  {journal} {Fluid Dynamics Research}\ }\textbf {\bibinfo {volume} {54}},\ \bibinfo {pages} {015505} (\bibinfo {year} {2022})}\BibitemShut {NoStop}%
\bibitem [{\citenamefont {Diamond}\ \emph {et~al.}(1995)\citenamefont {Diamond}, \citenamefont {Lebedev}, \citenamefont {Newman},\ and\ \citenamefont {Carreras}}]{diamond1995dynamics}%
  \BibitemOpen
  \bibfield  {author} {\bibinfo {author} {\bibfnamefont {P.}~\bibnamefont {Diamond}}, \bibinfo {author} {\bibfnamefont {V.}~\bibnamefont {Lebedev}}, \bibinfo {author} {\bibfnamefont {D.}~\bibnamefont {Newman}},\ and\ \bibinfo {author} {\bibfnamefont {B.}~\bibnamefont {Carreras}},\ }\bibfield  {title} {\enquote {\bibinfo {title} {Dynamics of spatiotemporally propagating transport barriers},}\ }\href@noop {} {\bibfield  {journal} {\bibinfo  {journal} {Physics of Plasmas}\ }\textbf {\bibinfo {volume} {2}},\ \bibinfo {pages} {3685--3695} (\bibinfo {year} {1995})}\BibitemShut {NoStop}%
\bibitem [{\citenamefont {Ashourvan}\ and\ \citenamefont {Diamond}(2017)}]{ashourvan2017emergence}%
  \BibitemOpen
  \bibfield  {author} {\bibinfo {author} {\bibfnamefont {A.}~\bibnamefont {Ashourvan}}\ and\ \bibinfo {author} {\bibfnamefont {P.~H.}\ \bibnamefont {Diamond}},\ }\bibfield  {title} {\enquote {\bibinfo {title} {On the emergence of macroscopic transport barriers from staircase structures},}\ }\href@noop {} {\bibfield  {journal} {\bibinfo  {journal} {Physics of Plasmas}\ }\textbf {\bibinfo {volume} {24}} (\bibinfo {year} {2017})}\BibitemShut {NoStop}%
\bibitem [{\citenamefont {Radko}(2013)}]{radko2013double}%
  \BibitemOpen
  \bibfield  {author} {\bibinfo {author} {\bibfnamefont {T.}~\bibnamefont {Radko}},\ }\href@noop {} {\emph {\bibinfo {title} {Double-diffusive convection}}}\ (\bibinfo  {publisher} {Cambridge University Press},\ \bibinfo {year} {2013})\BibitemShut {NoStop}%
\bibitem [{\citenamefont {Ouillon}\ \emph {et~al.}(2020)\citenamefont {Ouillon}, \citenamefont {Edel}, \citenamefont {Garaud},\ and\ \citenamefont {Meiburg}}]{ouillon2020settling}%
  \BibitemOpen
  \bibfield  {author} {\bibinfo {author} {\bibfnamefont {R.}~\bibnamefont {Ouillon}}, \bibinfo {author} {\bibfnamefont {P.}~\bibnamefont {Edel}}, \bibinfo {author} {\bibfnamefont {P.}~\bibnamefont {Garaud}},\ and\ \bibinfo {author} {\bibfnamefont {E.}~\bibnamefont {Meiburg}},\ }\bibfield  {title} {\enquote {\bibinfo {title} {Settling-driven large-scale instabilities in double-diffusive convection},}\ }\href@noop {} {\bibfield  {journal} {\bibinfo  {journal} {Journal of Fluid Mechanics}\ }\textbf {\bibinfo {volume} {901}},\ \bibinfo {pages} {A12} (\bibinfo {year} {2020})}\BibitemShut {NoStop}%
\bibitem [{\citenamefont {Mirjalili}, \citenamefont {Jain},\ and\ \citenamefont {Mani}(2022)}]{mirjalili2022computational}%
  \BibitemOpen
  \bibfield  {author} {\bibinfo {author} {\bibfnamefont {S.}~\bibnamefont {Mirjalili}}, \bibinfo {author} {\bibfnamefont {S.~S.}\ \bibnamefont {Jain}},\ and\ \bibinfo {author} {\bibfnamefont {A.}~\bibnamefont {Mani}},\ }\bibfield  {title} {\enquote {\bibinfo {title} {A computational model for interfacial heat and mass transfer in two-phase flows using a phase field method},}\ }\href@noop {} {\bibfield  {journal} {\bibinfo  {journal} {International Journal of Heat and Mass Transfer}\ }\textbf {\bibinfo {volume} {197}},\ \bibinfo {pages} {123326} (\bibinfo {year} {2022})}\BibitemShut {NoStop}%
\bibitem [{\citenamefont {Mirjalili}, \citenamefont {Jain},\ and\ \citenamefont {Dodd}(2017)}]{mirjalili2017interface}%
  \BibitemOpen
  \bibfield  {author} {\bibinfo {author} {\bibfnamefont {S.}~\bibnamefont {Mirjalili}}, \bibinfo {author} {\bibfnamefont {S.~S.}\ \bibnamefont {Jain}},\ and\ \bibinfo {author} {\bibfnamefont {M.}~\bibnamefont {Dodd}},\ }\bibfield  {title} {\enquote {\bibinfo {title} {Interface-capturing methods for two-phase flows: An overview and recent developments},}\ }\href@noop {} {\bibfield  {journal} {\bibinfo  {journal} {Center for Turbulence Research Annual Research Briefs}\ }\textbf {\bibinfo {volume} {2017}},\ \bibinfo {pages} {13} (\bibinfo {year} {2017})}\BibitemShut {NoStop}%
\bibitem [{\citenamefont {Mehdi-Nejad}, \citenamefont {Mostaghimi},\ and\ \citenamefont {Chandra}(2004)}]{mehdi2004modelling}%
  \BibitemOpen
  \bibfield  {author} {\bibinfo {author} {\bibfnamefont {V.}~\bibnamefont {Mehdi-Nejad}}, \bibinfo {author} {\bibfnamefont {J.}~\bibnamefont {Mostaghimi}},\ and\ \bibinfo {author} {\bibfnamefont {S.}~\bibnamefont {Chandra}},\ }\bibfield  {title} {\enquote {\bibinfo {title} {Modelling heat transfer in two-fluid interfacial flows},}\ }\href@noop {} {\bibfield  {journal} {\bibinfo  {journal} {International journal for numerical methods in engineering}\ }\textbf {\bibinfo {volume} {61}},\ \bibinfo {pages} {1028--1048} (\bibinfo {year} {2004})}\BibitemShut {NoStop}%
\end{thebibliography}%

\end{document}